\documentclass[%
 reprint,
superscriptaddress,
nofootinbib,
nobibnotes,
prb,
]{revtex4-2}
\usepackage{xcolor}
\usepackage{lineno}
\usepackage{graphicx}% Include figure files
\usepackage{hyperref}% add hypertext capabilities
\usepackage{amsmath}
\usepackage{color}
\usepackage{tabularx}
\usepackage{ulem}
\usepackage{footnote}
\usepackage{amssymb}
\begin{document}



\title{Collision Cascade-Driven Evolution of Vacancy Defects in Ni-Based Concentrated Solid-Solution Alloys}


\author{A. Aligayev}
\affiliation{%
NOMATEN Centre of Excellence, National Center for Nuclear Research, 
05-400 Swierk/Otwock, Poland
 }%
\affiliation{Science Island Branch of Graduate School, University of Science and Technology of China, Hefei, 230026, China}
%\email{Corresponding author: javier.dominguez@ncbj.gov.pl}
%\cortext[author] {Corresponding author: javier.dominguez@ncbj.gov.pl}
\author{M. Landeiro Dos Reis}
\affiliation{LaSIE UMR CNRS 7356, La Rochelle Université, Av. Michel Crépeau, 17042, La Rochelle Cedex 1, France}
\author{A. Chartier}
\affiliation{Université Paris-Saclay, CEA, Service de recherche en Corrosion et Comportement des Matériaux, 91191 Gif Sur Yvette, France}
\author{Q. Huang}
\affiliation{Science Island Branch of Graduate School, University of Science and Technology of China, Hefei, 230026, China}

\author{S. Papanikolaou}
\author{F. J. Dom\'inguez-Guti\'errez$^*$}
\affiliation{%
NOMATEN Centre of Excellence, National Center for Nuclear Research, 
05-400 Swierk/Otwock, Poland
 }%
\email{javier.dominguez@ncbj.gov.pl}

\def\thefootnote{$*$}
\footnotetext{Corresponding author\\F.J.D.G., e-mail: \url{javier.dominguez@ncbj.gov.pl}}
\def\thefootnote{\arabic{footnote}}

\begin{abstract}

Concentrated solid--solution alloys (CSAs) in single--phase form have recently garnered
considerable attention owing to their potential for exceptional irradiation resistance. 
This computational study delves into the intricate interplay of alloying elements on the generation, 
recombination, and evolution of irradiation-induced defects. 
Molecular dynamics simulations were conducted for collision cascades at room temperature, 
spanning a range of primary knock-on atom energies from 1 to 10 keV. 
The investigation encompasses a series of model crystals, progressing from pure Ni to binary
CSAs such as NiFe$_{20}$, NiFe, NiCr$_{20}$, and culminating in the more intricate NiFeCr$_{20}$
CSA.
%Our modeling framework takes into account a spectrum of vacancy defects, ranging from the least stable to the most stable in a face-centered cubic (FCC) material, including cubic, tetrahedral, octahedral, truncated octahedral, spherical, and stacking fault tetrahedral shapes. 
%Moreover, w
We observe that materials rich in chromium actively facilitate dislocation 
emissions and induce the nucleation of stacking fault tetrahedra in the proximity of nanovoids, 
owing to Shockley partial interactions.
This result is validated by molecular static simulations, which calculate the surface, 
vacancy, and defect formation energies. Among various shapes considered, the spherical 
void proves to be the most stable, followed by the truncated octahedron and octahedron shapes.
On the other hand, the tetrahedron cubic shape is identified as the most unstable, and stacking fault tetrahedra exhibit the highest formation energy.
%An intriguing observation emerged regarding the transformation from spherical vacancy defects to stacking fault tetrahedra (SFTs), a phenomenon intricately linked to the material's chemical composition.
%Specifically, only Cr-rich materials exhibited the nucleation of stair-rod dislocations, culminating in the formation of stable SFTs after collision cascades with PKA energies surpassing 5 keV. 
Notably, among the materials studied, NiCr$_{20}$ and NiFeCr$_{20}$ CSAs stood out as the sole alloys
capable of manifesting this mechanism, mainly observed at high impact energies.

\end{abstract}


\vspace{2pc}
%\noindent{\it Keywords}:

\maketitle

%\linenumbers
%########### sections
\section{Introduction}
\label{sec:intro}

Nickel and nickel-based concentrated solid solution alloys (CSAs) have been
extensively investigated for their potential applications
in harsh environments subjected to high irradiation doses 
\cite{lu2016enhancing,ullah2017effects,BWirth,C6CP05161H,LoyerProst1,MA2021116874} 
and could potentially serve as promising candidates for structural 
materials in nuclear applications, particularly in contexts where 
ensuring high radiation tolerance is a paramount consideration 
\cite{ULLAH201617,egami2014irradiation,PINTSUK20191300}.
Recent successes in the fabrication of CSAs have paved the
way for a new research direction aiming to substantially
enhance alloy performance 
\cite{lu2016enhancing,PhysRevLett.116.135504,JIN201665,kurpaska2022effects}. 
While historical alloy development focused on traditional
alloys with unique microstructural heterogeneity to mitigate
displacement damage 
\cite{zhang2016influence,PhysRevB.105.094117,PhysRevMaterials.7.025603,DELUIGI2021116951}, 
CSAs represent a shift, containing two to five or more
elements at high concentrations, sometimes in equal
or near--equal amounts.
The random arrangement of multiple elemental species on
a crystalline lattice results in atomic--level
elemental alternation, creating disordered local chemical
environments \cite{DELUIGI2021116951,wu2014recovery, gludovatz2014fracture}.
This intrinsic property leads to unique site--to--site
lattice distortions, creating complex energy landscapes
that affect defect migration 
\cite{senkov2010refractory, yang2018irradiation,LoyerProst1}.

Considering the challenging service conditions of 
next--generation nuclear power reactors, which involve 
intense radiation flux, higher operating temperatures, 
and high stress \cite{zinkle2013materials, allen2010materials},
CSAs emerge as promising candidate materials for
nuclear power applications \cite{yang2018irradiation,kurpaska2022effects}. 
To be viable, these alloys
must complement their superior mechanical properties
with high radiation resistance 
\cite{ullah2017effects,PhysRevLett.116.135504,PINTSUK20191300,MA2021116874}. 
In reactor environments, irradiation--induced point defect
formation, migration, and evolution are primary factors
influencing microstructural changes that impact the
performance of structural materials. 
Thus, controlling defect formation and migration in
structural materials becomes crucial for designing
materials with high radiation tolerance.
The phenomenon of irradiation hardening is governed by the
interactions between moving dislocations and 
irradiation induced defects occurring
at the atomic scale \cite{PhysRevB.105.094117,goryaeva2023compact}. 
For instance, at doses around 0.01 dpa, the observation
of softening becomes apparent in constant strain rate
traction tests for materials relevant to nuclear applications 
\cite{PhysRevMaterials.2.013604}. Upon the initiation of
plastic deformation, it is observed that the applied
stress initially decreases and subsequently stabilizes. 
This phenomenon is associated with the formation and 
propagation of shear bands, characterized by the absence
of any irradiation defect post--deformation \cite{PhysRevB.105.094117,PhysRevMaterials.7.025603}. 
The emergence of these bands strongly suggests that,
during their gliding, dislocations effectively eliminate
irradiation defects 
\cite{yang2018irradiation,PhysRevLett.116.135504}, 
which constitutes the primary focus of this work.

The kinetics and interaction of defects play a crucial role in
controlling microstructural evolution, significantly
impacting material properties over time 
\cite{arakawa2020quantum, chartier2019rearrangement,CHEN2021153124}. 
We thus investigate the influence of pre-existing defects on irradiation 
cascades, specifically examining common material defects such as vacancy 
clusters and stacking fault tetrahedra in elemental face-centered cubic 
(fcc) Ni, the two binary NiFe$_{20}$ and NiCr$_{20}$, and 
ternary NiFeCr$_{20}$ CSAs. We scrutinize several vacancy 
cluster shapes, from spherical to octahedron shape (in accordance with 
Wulff's predictions) to see if that influence the emergence of plasticity 
during irradiation cascade 
\cite{dos2020atomic,Boukouvala2021,PhysRevLett.99.135501,NIU2022104418,LIANG2022125997}. 
%The stability of vacancy clusters is influenced by their
Stacking fault tetrahedra (SFT), largely observed in this alloys are also considered. 
It is worth noticing that dislocation emission always begin from such pre-exisiting defects and that a clear influence of chromium is observed. 
It is interesting to note that in vicinity of voids Shockley dislocations interact to form SFT defects and stair-rod ones \cite{dos2020atomic}.
%stable until they reach a critical size, and the stable shape is contingent on the vacancy concentration.

%To explore the temporal progression of a displacement cascade, molecular 
%dynamics (MD) simulations are employed in a representative  metal \cite{nordlund2018improving,Boulle:nb5310,NORDLUND2018450}. 
%Typically initiated by the passage of a neutron or other high-energy 
%particle with MeV or greater energy, the cascade begins with the primary 
%knock-on atom— the first lattice atom to receive recoil energy. 
%Initially, as atoms undergo high excitation levels, many are displaced 
%from their lattice positions. However, as the cascade proceeds to 
%thermally equilibrate with its surroundings, nearly all atoms return to 
%their positions within the perfect lattice structure.

%{\color{red}{
%{\it{I am not sure that the paragraph below deserves to be here. It seems disconnected from the one above where the raison and the structure of the paper is presented.}}
%\newline

%}}

\section{Computational methods}
\label{sec:methods}
%\subsection{Interatomic potential}

The atomistic behavior of the samples was modeled using the 
Molecular Dynamics method as implemented in the Large-scale Atomic Molecular Massively Parallel 
Simulator (LAMMPS) \cite{THOMPSON2022108171}.% was employed for this purpose. 
 Interatomic potentials based on 
the embedded atom model (EAM)  developped by Bonny {\it et al} \cite{BONNY201742,Bonny_2013} 
were employed to describe atom--to--atom interactions. 
%
%The choice of the potential stems notably from its ability to accurately 
%reproduce key physical properties for this study, including elastic 
%constants and the Generalized Stacking Fault (GSF) energy. %(see Table 
%\ref{tab:physical_properties}, and Fig. \ref{fig:GSF}). 
%The methodology for computing these properties is detailed in the supplementary materials.
%
%\newline
%{\color{blue}{
%{\it{I do not understand how it is possible to generate a random distribution with a cell containing 4 atoms only (0.352 nm cell is the conventional cell, and thus containing 4 atoms for a fcc lattice). So, I do believe that there is a mistake below. So, you should indicate the exact number of atoms of the supercell used to generate the alloys.}}
%\newline
For computations, we initiated a numerical cell for pure Ni with 
[001] crystal orientations using the Atomsk numerical tool 
\cite{HIREL2015212}, employing a lattice constant of 0.352 nm. 
This initial configuration was then duplicated to create a 
supercell with varying volumes for different calculations in this 
study. Subsequently, alloy configurations were generated by 
introducing random substitutions in the supercell. 
This involved introducing 50\% and 20\% Fe 
for NiFe and NiFe$_{20}$ CSAs, 20\% Cr for NiCr$_{20}$, and 40\% 
Fe with 20\% Cr for NiFeCr$_{20}$ CSAs.
To account for randomness, computations were conducted
across several different random alloy configurations. 
%}}

The cells were relaxed using FIRE minimization algorithm
until the force reach $1 \times 10^{-3}$ meV/\AA{} \cite{GUENOLE2020109584} where the samples find their 
lowest energy structure.
It is noteworthy that the optimization
process for the CSAs' geometry was aimed at reaching 
the nearest local minimum of the energy structure ensuring
that the change in energy between successive 
iterations and the most recent energy 
magnitude remains below 10$^{-5}$. 
Additionally, the global force vector length of all
atoms is maintained at less than or equal
to $10^{-8}$ eV/\AA{}.

%\subsection{Elastic constant}
%
%The elastic constants were calculated using the stress-strain tensors
%in the limit of infinitesimal deformation method. In this approach, the lattice
%vectors of the simulated system were strained incrementally, and the
%corresponding stress tensors were computed using the virial stress formulation 
%\cite{PhysRevB.69.134103}. The resulting stress--strain data were then fitted
%to linear regressions to extract the elastic constants. 
%Here, the elastic tensor, expressed as a \(6 \times 6\) matrix using the
%Voigt concept (\(C_{ij}\) instead of \(C_{ijkl}\)), allows for the calculation
%of elastic constants by applying elementary deformations to six distinct strain
%components and measuring the resulting changes in six stress components. 
%Thus, we build simulation cells to study the solid--solution alloys and
%the pure Ni crystal in the $\{001\}$ orientation with $N_{at} \simeq 863 000$ atoms, 
%arranged within dimensions $(d_x, d_y, d_z)$ = (21, 21, 21) nm;  
%a strain of \(10^{-6}\) was applied to induce deformation
%in the simulation box. Consequently, by analyzing the stress-strain
%relationship, the elastic constants can be determined.

\subsection{Molecular Static simulations}

Point defects like vacancies where a missing atom at a lattice point 
can impact a CSAs' chemical properties and mechanical behavior during 
collision cascades are analyzed by computing the formation energy ($E_v$) 
of a vacancy, crucially determining the energy needed to break bonds
and remove an atom from the crystal. 
To achieve this, we employ a computational cell containing
$N_v =$ 4000 atoms, with a lateral dimension of 3.52 nm. 
The vacancy energy is then calculated using the following method:
\begin{equation}
    E_v = E_v^f - \frac{N_v-1}{N_v}E_v^i, 
\end{equation}
where $E_v^f$ and $E_v^i$ are the total energy with the vacancy and the 
pristine one, respectively.

%\subsection{Surface energy}
%We used cells with free surfaces in $Z$ direction and periodic boundary conditions in $X$ and $Y$ directions, following:
%\begin{itemize}
%    \item $X=[1 0 0]$, $Y=[0 1 0]$, $Z=[001]$ for (100) surface
%    \item $X=[1 0 0]$, $Y=[0 1 \Bar{1}]$, $Z=[011]$ for (110) surface
%    \item $X=[1 \Bar{1} 0]$, $Y=[1 1 \Bar{2}]$, $Z=[111]$ for (111) surface
%\end{itemize}
%Dimensions $(d_x, d_y, d_z) = (2.5, 2.5, 9)$ nm were utilized. 
Then we checked 2D defects, surface and stacking fault.
Surface energy of the $(hkl)$ directions were computed following:
\begin{equation}
   \Gamma_{hkl} = \frac{E_{\rm surf}-E_0}{2S}
   \label{eq:surface_energy}
\end{equation}
with $E_{\rm surf}$ the energy of the cell, $E_0$ the energy of 
the pristine cell and $S$ the surface area.

%\subsection{Generalized Stacking Fault}
%For the computation of the Generalized Stacking Fault (GSF) energy, a cell with the orientation of $X=[1 \Bar{1} 0]$, $Y=[1 1 \Bar{2}]$, and $Z=[111]$ and dimensions $(d_x, d_y, d_z) = (2.5, 2.5, 9)$ nm were utilized.
%Periodic boundary conditions were used in the $X$ and $Y$ directions, while free surfaces were allowed in the $Z$ direction. 
%Solute atoms were uniformly and randomly distributed across --> already said above
10 different samples following the same atomic percentages for the CSAs were employed to ensure a reliable average of the Generalized Stacking Fault (GSF) energy, denoted as $\gamma_{\rm GSF}$.
The stacking fault was introduced by rigidly shifting the
half upper region of the cell in the $X$ and $Y$
directions. 
%Subsequent relaxation of the cell occurred until the interatomic force reached 0.1 meV/\AA.
 
The energy of the cell, denoted as $E_{\rm GSF}$, was then 
computed based on the displacement of the cell in the $Y$ direction.% (Fig. \ref{fig:GSF}).
\begin{equation}
    \gamma_{GSF} = \frac{E_{\rm GSF}-E_0}{A}
\end{equation}
with $E_0$ the energy of the pristine cell without the 
stacking fault %, $i.e$. without any shift, 
and $A$ is the GSF area. 
%The simulation was repeated ten times with 
%varying randomizations, providing 
%an averaged set of elastic constants.
%The cells were relaxed using FIRE minimization algorithm until the force reach 1x10$^{-3}$ meV/\AA. 

%\subsection{Defects}

%Our investigation focused on various alloys, namely NiFe, NiFe$_{20}$, 
%NiCr$_{20}$, and NiFeCr$_{20}$ which were compared to pure Ni case.
%Solute atoms were randomly introduced by substituting pure crystal
%Ni atoms. 
\par
In third part we studied 3D defects. It is well--established that there are various defects in fcc metals  \cite{eyre1973transmission}. 
Particularly, vacancy clusters and stacking fault tetrahedra (SFT) 
are prominently observed defects in these materials
\cite{chakraborty2017cr}. In this work, we decided to evaluate the 
impact of such defects (voids and SFT) on the irradiation cascade 
evolution. The procedure to introduce the defects is described in the following subsections.

%\subsubsection{Voids and Stacking fault tetrahedra (SFT)}

%{\color{red}{
%{\it I think that the paragraph concerning the shapes of the truncated octahedra using the Wulff theory and presented in the result part must be here. So I copy paste the paragraph from the results.tex part below. Since it is just a copy paste, might have to be rewritten}
%\newline
Vacancy clusters were built by removing atoms within the geometric region defining the defect. We built voids of spherical, octahedral, and truncated octahedral shapes (as predicted by the Wulff theory using surface energies presented in the results section), as well as tetrahedral and cubic shapes.
%\textcolor{blue}{ 
The Wulff theory, which relies on surface energy calculations
across primary surface planes %(see Table \ref{tab:surface_energy}), 
enables us to anticipate the stable configuration of vacancy 
clusters. This theory is in good agreement with atomistic 
simulations for sufficiently large clusters, where the faces
of the cluster can be regarded as free surfaces. 
However, the theory encounters limitations when dealing
with small clusters where the ratio of edge to face is high.
Therefore, we initially investigated the stability of defects 
prior to irradiation cascade by employing Wulff theory %{(Fig. \ref{eq:formation_energy})} 
alongside the analysis of formation 
energy results %(see method, Sec. \ref{Sec:Formation_energy}).
by using a Python package for Wulff construction \cite{marks1983modified,wul1901frage} to determine the Wulff shape using $\Gamma_{100}$, $\Gamma_{100}$ and $\Gamma_{111}$.
%given Table \ref{tab:surface_energy}.
%}
%}}
%\newline

%\subsubsection{}
The SFT was implemented employing the approach delineated in Ref. 
\cite{kadoyoshi2007molecular}. Following the same methodology as for voids, we built a triangular vacancy plate that relaxed into SFT defects with stair-rod dislocations, due to the low stacking fault energy of the CSAs, as predicted by the Hirsch and Silcox mechanism \cite{silcox1959direct}.

%\subsubsection{Formation energy of defects \label{Sec:Formation_energy}}
To assess the stability of these defects prior to irradiation we computed the mean formation energy of each defects. For these peculiar simulations, the cells involved $N_{at} \simeq 863 000$ atoms, arranged within dimensions $(d_x, d_y, d_z)$ = (21, 21, 21) nm. The defects were introduced as described before, and the mean formation energy per vacancy, $E_f$, of these defects was computed as follows:
\begin{equation}
    E_f = \frac{1}{N_{at} - N_d} \left( E_{d} - \frac{N_d}{N_{at}} E_0 \right)
    \label{eq:formation_energy}
\end{equation}
where $E_d$ is the energy of the simulation cell with the defect, 
containing $N_d$ atoms and $N_{at}-N_d$ vacancy and $E_0$ is the 
energy of the simulation cell of the pristine CSA, containing 
$N_{at}$ atoms. We chose to rationalize the formation energy based on the number of vacancies, as defects of the same size can have different numbers of vacancies due to their unique shapes.
%The cells were relaxed using FIRE minimization algorithm until the force reach 1x10$^{-3}$ meV/\AA.  

%\subsection{Defect identification}

%%%%%%%%%%%%%%%%%%%%%%%%%%%%%%%%%%%%%%%%%%%%%%%%%%%%%%%
\subsection{Collision cascades}

%The selection of the simulation cell size for a single collision cascade is based on ensuring the inclusion of all atoms within a radius $r_0$, which defines the minimal size of the cell with 50--100 atoms (typically around 3 \AA{}), as discussed in  
%\cite{NORDLUND1995448,Nordlund2018_Nature}. 
%However, this small cell size is inadequate to encompass the entire trajectory of an implanted ion in the keV energy range, as such ions may traverse larger distances within the implanted sample. Therefore, a mechanism is required to ensure that the recoiling atom is always surrounded by lattice atoms within a larger cell \cite{nordlund2018improving,Boulle:nb5310,NORDLUND2018450}.
The simulation cell size for a single collision cascade is selected to 
include all atoms within a radius $r_0$ ($\sim$ 3 \AA{}), as 
discussed in \cite{NORDLUND1995448,Nordlund2018_Nature}. However, this size 
is insufficient for the trajectory of an implanted ion in the keV range. To 
address this, a larger cell ensures the recoiling atom is always surrounded 
by lattice atoms \cite{nordlund2018improving,Boulle:nb5310,NORDLUND2018450}.
A pure fcc Ni sample with $(d_x,d_y,d_z)$=(9.85,10.20,10.56) nm is initially 
created. After energy minimization, a 100 ps equilibration is conducted at 
300 K with a time constant of 100 fs \cite{Javier2022,Dominguez-Gutierrez_2021}. 
%The sample is equilibrated at $T = 300$ K for 2 ns using an 
%isobaric--isothermal ensemble, maintaining an external pressure of 0 GPa\cite{kurpaska2022effects}.
%This process continues until achieving a 
%homogeneous temperature and pressure profile.

\begin{figure}[t!]
    \centering
    \includegraphics[width=0.48\textwidth]{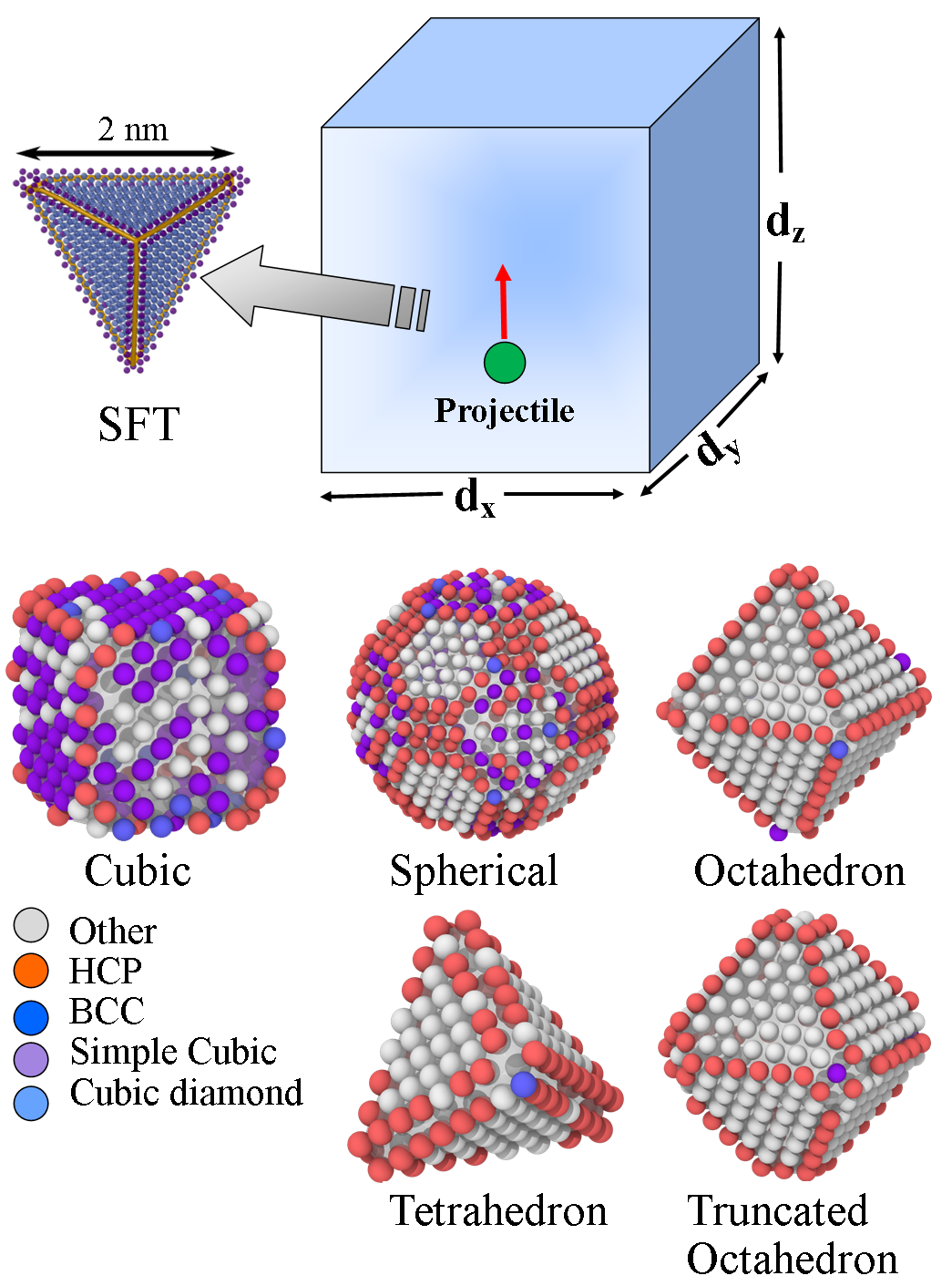}
    \caption{(Color online) Schematics of the numerical cell utilized for conducting collision cascades, encompassing various defects such as cubic, spherical, octahedron, tetrahedron, truncated octahedron, and stacking fault tetrahedra. Initiated by assigning kinetic energy to a randomly selected atom-projectile, defects are characterized by identifying their atomic structure for enhanced visualization.}
    \label{fig:sampleFigure}
\end{figure}

%To explore the influence of Fe and Cr atoms on defect production during 
%collision cascades within the Ni matrix, the preparation of the alloys
%mirrored that of the pure Ni sample. 
%We then introduce defects into the simulation cell following the explanation provided in the previous sections for all the CSAs. 
As illustrated in Fig. \ref{fig:sampleFigure}, we considered two kind of 
defects: nanovoids and SFT, commonly found in
this fcc metals \cite{eyre1973transmission}. 
The size of the defects is set to 2.0 nm.
Regarding nanovoids, we analyzed the impact of their shape and performed tests on different 
geometries, encompassing spherical, octahedral, %(with or without truncation on ${100}$ planes) 
%consistent with the Wulff shape prediction \cite{dos2020atomic}),
tetrahedral, and cubic shape.

%\subsection{collision cascades}
%The initiation of a MD simulation for performing collision cascades
%involves the random selection
%of a Fe, Ni, or Cr atom located at the center volume of the 
%numerical sample, with kinetic energy (KE) assigned in the
%range of 1--10 keV to the Primary Knock-On Atom (PKA). 
%To demonstrate that radiation damage production is independent
%of stochastic anomalies, collision cascade simulations are 
%conducted with 50 randomly directed energetic PKAs for 
%various energies (1, 2, 5, 8, and 10 keV).
%Then, the Velocity Verlet integration algorithm is applied to model the 
%collision cascade for 6 ps, followed by an additional relaxation
%time of 4 ps, as depicted in Fig. \ref{fig:sampleFigure}.  
%A Nosé–Hoover thermostat was employed on a 6-\AA{} thick shell 
%at the simulation cell boundary to gradually cool the cell back
%to its initial temperature. This approach emulates the behavior of the 
%significantly larger bulk material encompassing the cascade region. 
%Notably, pressure control was not applied during the cascade simulations.
%\cite{chartier2019rearrangement,KOSKENNIEMI2023154325,Dominguez-Gutierrez_2021}
MD simulations for collision cascades start with randomly selecting
a Fe, Ni, or Cr atom at the center of the numerical sample. 
The Primary Knock-On Atom (PKA) receives kinetic energy (1--10 keV),
and 50 simulations are conducted with arbitrary velocity direction. 
The Velocity Verlet integration algorithm models the cascade
for 6 ps, followed by 4 ps relaxation, as shown in Fig. 
\ref{fig:sampleFigure}. A Nosé–Hoover thermostat cools the cell
back to the initial temperature, simulating the bulk material's
behavior. Pressure control is not applied during cascade simulations 
\cite{chartier2019rearrangement,KOSKENNIEMI2023154325,Dominguez-Gutierrez_2021}.

\subsection{Analysis of collision cascades}
\label{subsec:analysisCascades}

The analysis of the formation of defects and dislocations during collision cascades 
is performed by computing the dislocation length as a function of simulation time 
for all the samples using OVITO software \cite{ovito}. 
We utilized the polyhedral template matching (PTM) to identify different atomic 
structures and the 
Dislocation Extraction Algorithm (DXA)~\cite{Stukowski_2010} which
extracts dislocation structure and content from 
atomistic microstructures. 
Thus, we categorized the atomic structures as: BCC, FCC, HCP, icosahedral, simple 
cubic, and cubic diamond; the dislocations into several dislocation
types according to their Burgers vectors as: ½$\langle110\rangle$ 
(Perfect), 1/6$\langle112\rangle$ (Shockley), 1/6$\langle110\rangle$ 
(Stair--rod), 1/3$\langle100\rangle$ (Hirth), 1/3$\rangle111\langle$ 
(Frank). Then the dislocation density is obtained as
\begin{equation}
    \rho(t) = L(t)/V_c
\end{equation}
where $L(t)$ is the dislocation length of different types and $V_c$ is 
the cell volume. In addition, the vacancies and voids are identified by 
the Delaunay tessellation to partition space into tetrahedral elements that 
are categorized as either solid or empty through the computation of an alpha parameter 
implemented into the surface mesh tool in OVITO \cite{stukowski2014computational}.

%\begin{table*}[t!]
%\caption{.} 
%    \centering
%    \begin{tabular}{c|ccccc}
%         &  Ni & NiFe$_{20}$ & NiFe & NiFeCr$_{20}$ & NiCr$_{20}$ \\
%         \hline
%         \hline 
%        $\Gamma_{100}$ (meV/\AA²) & 48.4 & 70 $\pm$ 2 & 93$\pm$ 2& 95 $\pm$ 2 & 58 $\pm$ 2\\
%        $\Gamma_{110}$ (meV/\AA²) &  64.5 & 85 $\pm$ 2 & 106 $\pm$ 2& 108 $\pm$ 2 & 74 $\pm$ 2\\
%        $\Gamma_{111}$ (meV/\AA²) &  43.0 & 65 $\pm$ 2 & 89 $\pm$ 2& 89 $\pm$ 2 & 53 $\pm$ 2\\
%        
%    \end{tabular}
%    
%\end{table*}

\section{Results}
\label{sec:results}

It is noteworthy that Cr and Fe have the capability to modify the 
Stacking Fault Energy (SFE), a critical parameter in understanding 
dislocation behavior and the recovery process 
\cite{zaddach2013mechanical,schramm1976stacking,carter1977stacking,
ullah2020electron}. Materials with low SFE 
promote the formation of deformation twins, partial dislocations 
with a wide stacking fault ribbon, and a high stacking fault 
density \cite{youssef2011effect, wu2013effect}. This stacking 
fault acts as a barrier against cross-slip or climb mechanisms, 
leading to slower recovery and the development of materials with 
high strength and good ductility. In contrast, materials with high 
SFE exhibit more rapid cross-slip and climb, 
resulting in a higher recovery rate \cite{gong2013simultaneously, 
wu2014recovery}.
The modeling of radiation defects production has to accounting
for the transfer of kinetic energy from high-energy incident 
particles to the lattice atoms in the surface sample. 

Vacancy formation energies were calculated in good agreement with 
reported data by S. Zhao et al. \cite{C6CP05161H}, with values of 1.48 eV for pure Ni, 
1.50 eV $\pm$ 0.2 eV for NiFe$_{20}$ CSA while removing a Fe or a
Ni atom; 1.70 $\pm$ 0.2 eV for equiatomic binay NiFe CSA; and 
0.97 $\pm$ 0.2 eV for NiCr$_{20}$ CSA.
Finally, our calculation for NiFeCr$_{20}$ CSA is 1.89 $\pm$ 0.25 eV 
in good agreement with reported data by Manzoor et al. \cite{10.3389/fmats.2021.673574}

\subsection{Surface and Generalized Stacking Fault energy (GSF)}
\begin{table}[b!]
\caption{Surface energy $\Gamma_{hkl}$ of the Ni, NiFe and NiFeCr systems computed using Eq. \ref{eq:surface_energy} in (meV/\AA{}$^2$)}
    \centering
    \begin{tabular}{l|ccc}
         & $\Gamma_{100}$  &  $\Gamma_{110}$ & $\Gamma_{111}$  \\
         \hline \hline
       Ni          & 48.4     &   64.5          & 43.0 \\
       NiFe$_{20}$ & 70 $\pm$ 2 & 85 $\pm$ 2 & 65 $\pm$ 2 \\
       NiFe        & 93 $\pm$ 2 & 106 $\pm$ 2 & 89 $\pm$ 2 \\
       NiCr$_{20}$ & 58 $\pm$ 2 & 74 $\pm$ 2 & 53 $\pm$ 2 \\
    NiFeCr$_{20}$  & 95 $\pm$ 2 & 74 $\pm$ 2 & 53 $\pm$ 2
    \end{tabular}
    \label{tab:surface_energy}
    \label{tab:my_label}
\end{table}

%The FCC structure is characterized by three independent elastic constants:
%C$_{11}$, C$_{12}$, and C$_{44}$, determined through lattice strain, as presented
%in Table \ref{tab:physical_properties}.

%\begin{table}[b!]
%    \centering
%    \caption{Elastic constant computed with EAM potential \cite{BONNY201742,Bonny_2013} compared to 
%    literature, we include statistical errors due to material's randomness. 
%    %\Marie{There is no error bars on the results for elastic constant link to the randomness of the alloys ? And comparison may be to some DFT or experimental data }
%    }
%    \begin{tabular}{c|cccc}
%         & $C_{11}$ (GPa) & Exp. &  $C_{12}$ (GPa) & $C_{44}$ (GPa)   \\
%         \hline
%         \hline
%       Ni            & 247         & - & 147 & 125 \\
%       NiFe$_{20}$   & 298 $\pm$ 1 & 235 & 162 $\pm$ 1 & 145 \\
%       NiFe          & 314 $\pm$ 2 & & 165 $\pm$ 1 & 158 \\
%       NiCr$_{20}$   & 197 $\pm$ 1 & & 132 $\pm$ 1 & 107 \\
%       NiFeCr$_{20}$ & 261 $\pm$ 2 &  & 158 $\pm$ 1 & 142 $\pm$ 1 \\
%    \end{tabular}
%    \label{tab:physical_properties}
%\end{table}
In order to provide information about the predefined 
defects in the alloys, we compute the surface energy of the 
pristine CSAs by using Eq. 1 that are displayed in Tab. 
\ref{tab:surface_energy}. 
Fe significantly increases the surface energy across all three surfaces, 
whereas the impact of chromium is comparatively less noticeable. However,
neither Fe nor Cr altered the stability hierarchy among the three 
primary planes.

\begin{figure}[b!]
    \centering
    \includegraphics[scale = 0.5 ]{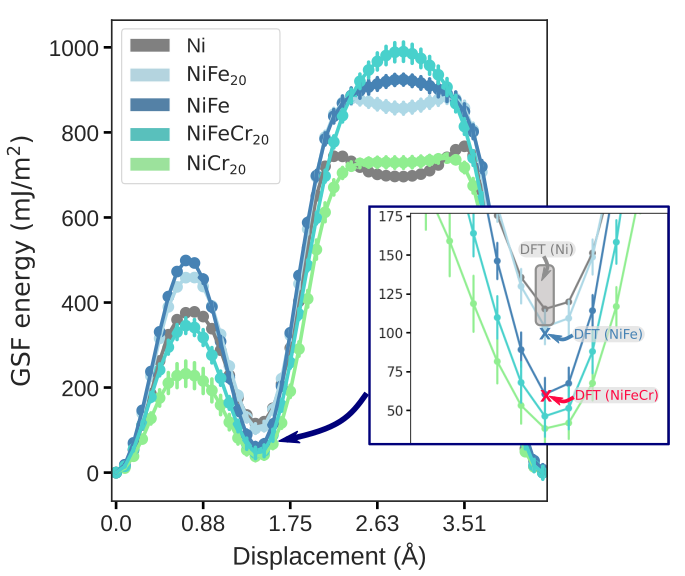}
    \caption{ Generalize the Stacking Fault energy across diverse alloy compositions compared to DFT \cite{zhao2017stacking,zaddach2013mechanical}. Uncertainties arise from the random distribution of solute atoms.}
    \label{fig:GSF}
\end{figure}

Results for the GSF energies of the pristine CSAs are shown in
Fig. \ref{fig:GSF} for all the systems. 
The intrinsic stacking fault energy is the metastable 
point of the GSF energy (see magnification on Fig. \ref{fig:GSF}). 
 We show a consistent decrease
 in the intrinsic stacking energy (see magnification in Fig. 
 \ref{fig:GSF}) with increasing the concentration of Fe or Cr 
 solutes. 
 This reduction is particularly prominent with Cr solutes, 
 showcasing a substantial
 drop from 116 mJ/m$^2$ for pure Ni to 38 mJ/m$^2$ when Ni 
 incorporates only 20\% Cr. 
 Conversely, the impact of Fe solutes is more subdued at 104 
 mJ/m$^2$ and escalates with an increasing number of
 Fe solutes, reaching 60 mJ/m$^2$ for NiFe alloys. 

 Intriguingly, both Fe and Cr solutes independently contribute
 to a decrease in intrinsic SFE. However, in the ternary
 alloy NiFeCr$_{20}$, the introduction of Fe solutes induces a noteworthy
 increase, up to 47 mJ/m$^2$, in intrinsic SFE compared
 to NiCr$_{20}$ alloys.
Finally, to assess validity of the potential, our results were compared to those provided in Ref. \cite{zhao2017stacking, zaddach2013mechanical}, for which more precise atomistic simulations were performed, employing  Density
 Functional Theory method (DFT) (Fig. \ref{fig:GSF}) reaching 
 a good agreement.

\subsection{Stability of defects before irradiation cascade}

\begin{figure}[t!]
    \centering
    \includegraphics[width=0.48\textwidth]{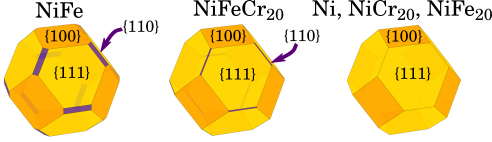}
    \caption{Wulff shape predictions computed based on the surface energy of alloys (Table \ref{tab:surface_energy})}
    \label{fig:wulff}
\end{figure}

%\sout{The Wulff theory, which relies on surface energy calculations
%across primary surface planes (see Table \ref{tab:surface_energy}), 
%enables us to anticipate the stable configuration of vacancy 
%clusters. This theory is in good agreement with atomistic 
%simulations for sufficiently large clusters, where the faces
%of the cluster can be regarded as free surfaces. 
%However, the theory encounters limitations when dealing
%with small clusters where the ratio of edge to face is high.
%Therefore, we initially investigated the stability of defects 
%prior to irradiation cascade by employing Wulff theory (Fig. 
%\ref{eq:formation_energy}) alongside the analysis of formation 
%energy results (see method, Sec. \ref{Sec:Formation_energy}).
%
%\subsection{Wulff Theory}
%
%We used a Python package for Wulff construction \+cite{marks1983modified,wul1901frage} to determine the Wulff shape using $\Gamma_{100}$, $\Gamma_{100}$ and $\Gamma_{111}$ given Table \ref{tab:surface_energy}.}
The predicted shape are depicted Fig. \ref{fig:wulff}. Truncated octahedron are generally the preferred shape for vacancy clusters across all alloys, albeit with slight variations in the proportion of $\{110\}$ planes.

%%%%%%%%%%%%%%%%%%%%%%%%%%%%
\subsubsection{Formation energy of defects}

\begin{figure}[b!]
    \centering
    \includegraphics[scale=0.5]{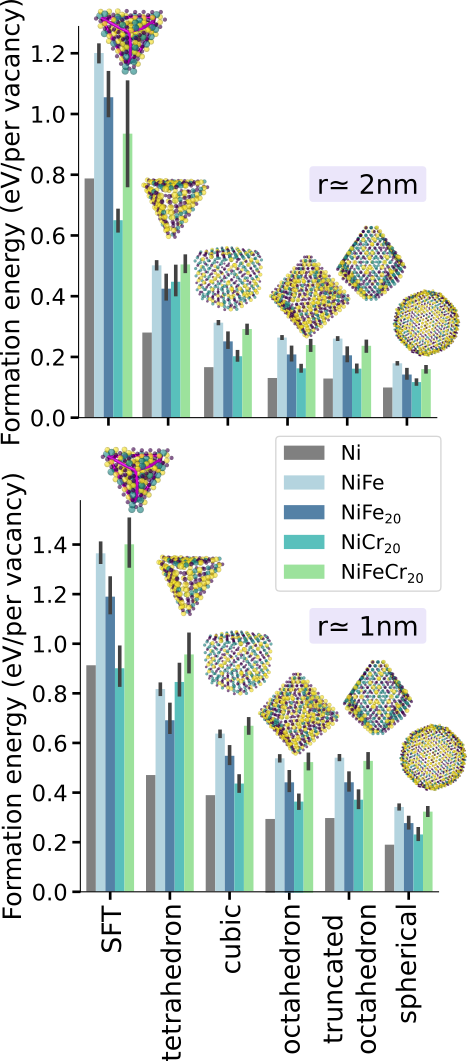}
    \caption{Mean formation energy per vacancy of defects according to their shape and their size ($r$) for the different alloys. The results are averages derived from 10 different atomic configurations of alloys, and the associated error bars represent the standard deviations. }
    \label{fig:formation_energy}
\end{figure}

The results are shown Fig. \ref{fig:formation_energy}, for all alloys and 
all defects. The results are averages derived from 10 different atomic 
configurations of alloys, and the associated error bars represent the 
standard deviations. When comparing the alloys, an observation is that 
the presence of Fe atoms leads to an increase in the defect 
formation energy. The highest formation energies are obtained in CSAs
with the highest concentration of Fe, such as NiFe alloys, 
while the lowest formation energy is observed in binary NiCr CSAs.
A lower formation energy indicates that defects form easily. 
Consequently, our findings suggest that the presence of iron
may hinder the formation of vacancy clusters compared to both
pure Ni and NiCr$_{20}$. According to our simulations spherical
shape is the most stable, following by truncated octahedron
and octahedron shape. The most unstable shape is tetrahedron cubic.
SFT exhibit the highest formation energy; however, a direct
comparison with voids is challenging because SFTs
are composed of dislocation stair rods, not solely vacancies.

\subsection{Collision cascade}

\begin{figure}[t!]
    \centering
    \includegraphics[width=0.45\textwidth]{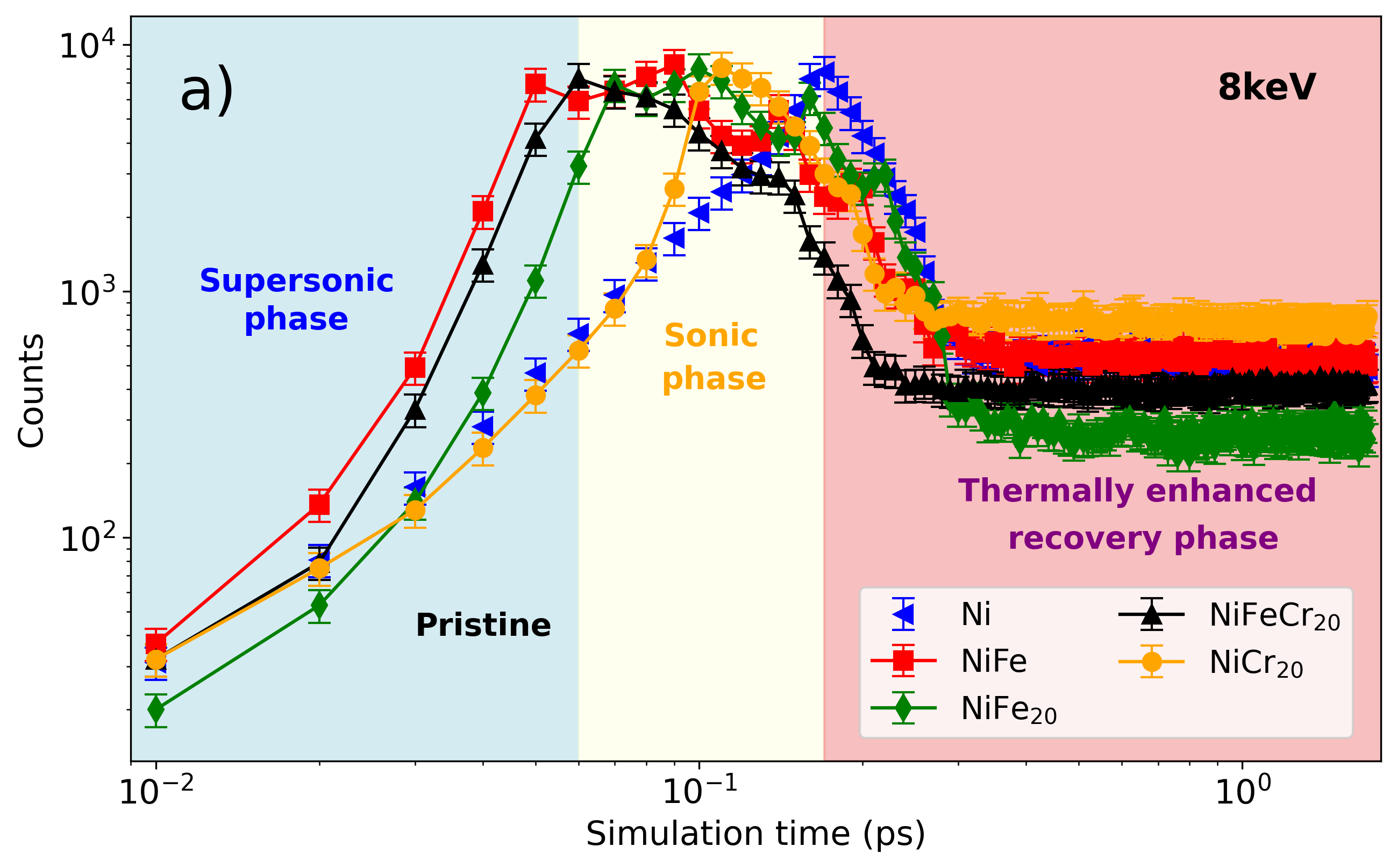}
    \includegraphics[width=0.45\textwidth]{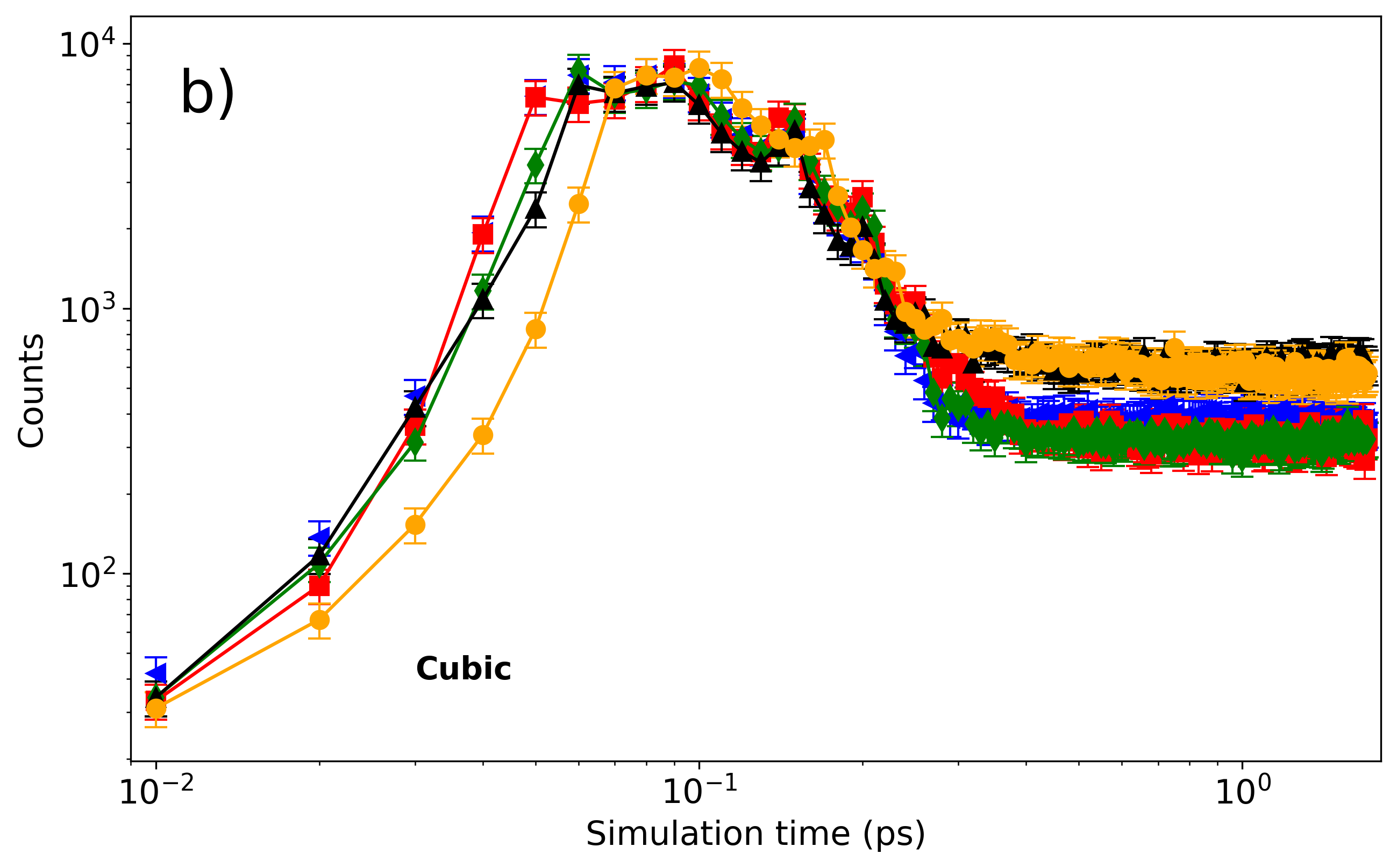}
    \includegraphics[width=0.45\textwidth]{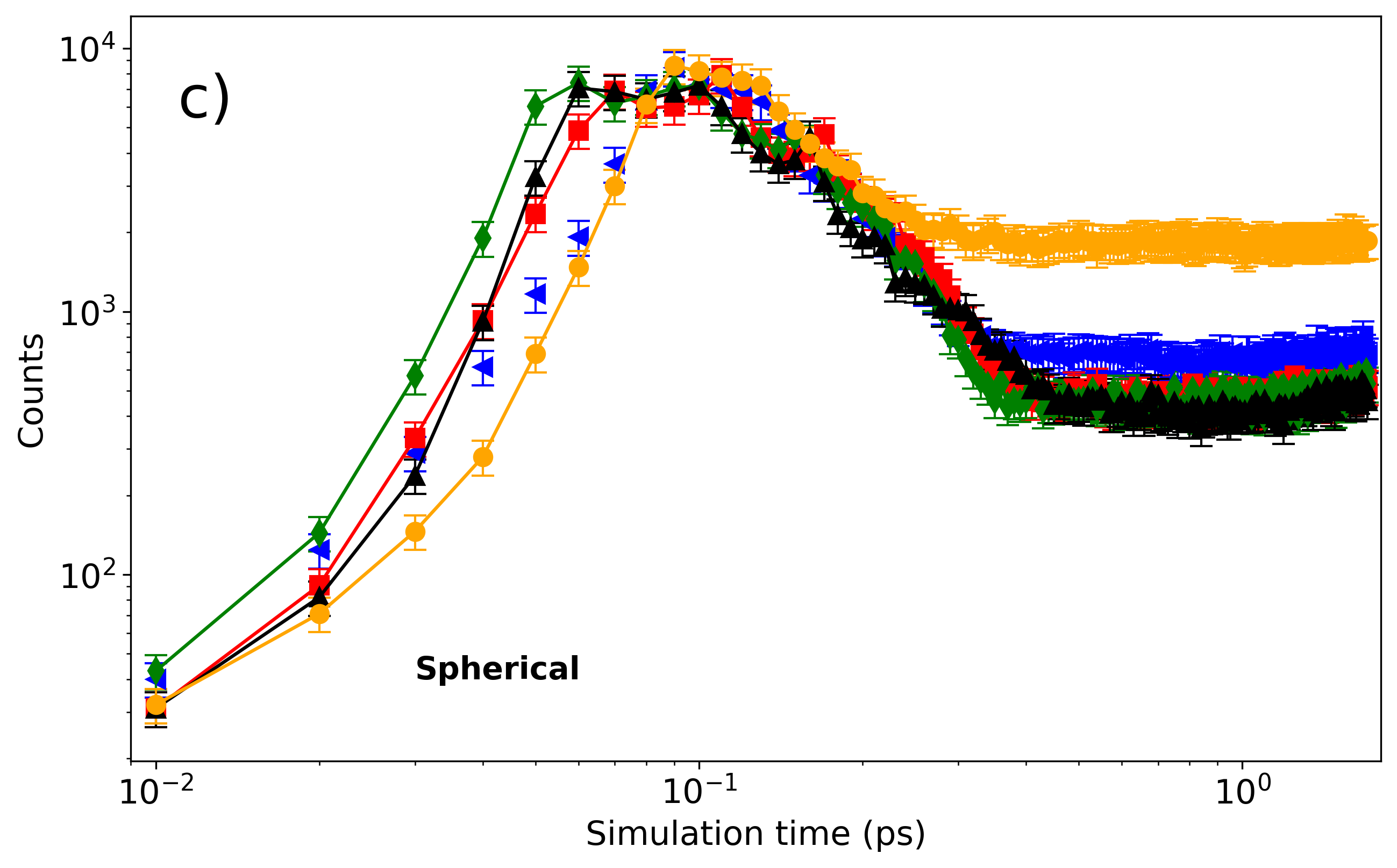}
   \caption{(Color online) Typical time profiles depicting the number
   of defects during 8 keV cascades in Ni and its solid solution
   alloys for a defect--free sample in a), a cubic defect in b), and an 
   spherical void in c). 
   The three cascade stages—supersonic, sonic, and thermally 
   enhanced recovery—are illustrated by the background colors in (a).}
    \label{fig:Fig1R}
\end{figure}

Fig. \ref{fig:Fig1R} presents the outcomes of collision cascades 
involving pristine Ni and its CSAs for a defect--free sample
in a), a cubic defect in b), and an spherical void in c). 
The temporal evolution of defects considered as atoms with a different 
structure than FCC during collision cascades is characterized by the phases
of supersonic, 
sonic, and thermal recovery with respect to pure Ni, in good agreement
with reported results \cite{fabris2018theoretical}. 
Additionally, a notable effect observed in the solid solution alloys, 
compared to Ni, is the duration of the supersonic and sonic phases 
leading to the recovery phase.
The influence of Fe and/or Cr atoms in the Ni sample is evident in
the rapid production of defects by equiatomic NiFe and
NiFeCr$_{20}$ alloys, where the inclusion of cubic and spherical defects 
modified the behaviour of the material changing the recovery time after 
collision cascade. The introduction of Cr in the Ni sample
accelerates defect production after 0.1 ps, attributed to Ni--Cr
and Cr--Cr interactions and the large decrease of the intrinsic 
stacking fault energy (Fig. \ref{fig:GSF}), noticing a high defect production 
for the spherical void case.
In the presence of Fe in NiFe$_{50,20}$ and NiFeCr$_{20}$ CSAs,
the material requires more time to recover. 
Conversely, the presence of Cr does not significantly impact
these phases. These results are used as reference for our work.

To explore the impact of preexisting defects on point defect 
formation 
during collision cascades, MD simulations were conducted across 
various 
PKA values and materials. In Fig \ref{fig:Fig1R}, we 
present results at 8 keV PKA, showcasing defect count profiles 
over simulation time for the most unstable defect vacancy
(cubic) in b) and the most stable defect vacancy (spherical)
in c), which can be considered as a void in the CSAs 
\cite{PhysRevMaterials.4.103603}; however, these defects 
can respond variably to irradiation. A comparison with 
the pristine case reveals observable changes in the defect 
production mechanism across all MD simulations. 
Additional results are available in the supplementary material.
Distinct effects of Cr on defect production in Ni and
NiFe alloys were noted, leading to an increase in
defect counts. Conversely, a small percentage of Fe
in the Ni matrix resulted in reduced defect counts after 
the collision cascade in the pristine case. 
Intriguingly, when 50\% of Fe is present in Ni, a drastic 
reduction in defects is observed for both 
cubic and SFT cases. 
This could potentially be an artifact of the Bonny potential
\cite{beland2017accurate}.

\begin{figure}[t!]
    \centering
    \includegraphics[width=0.40\textwidth]{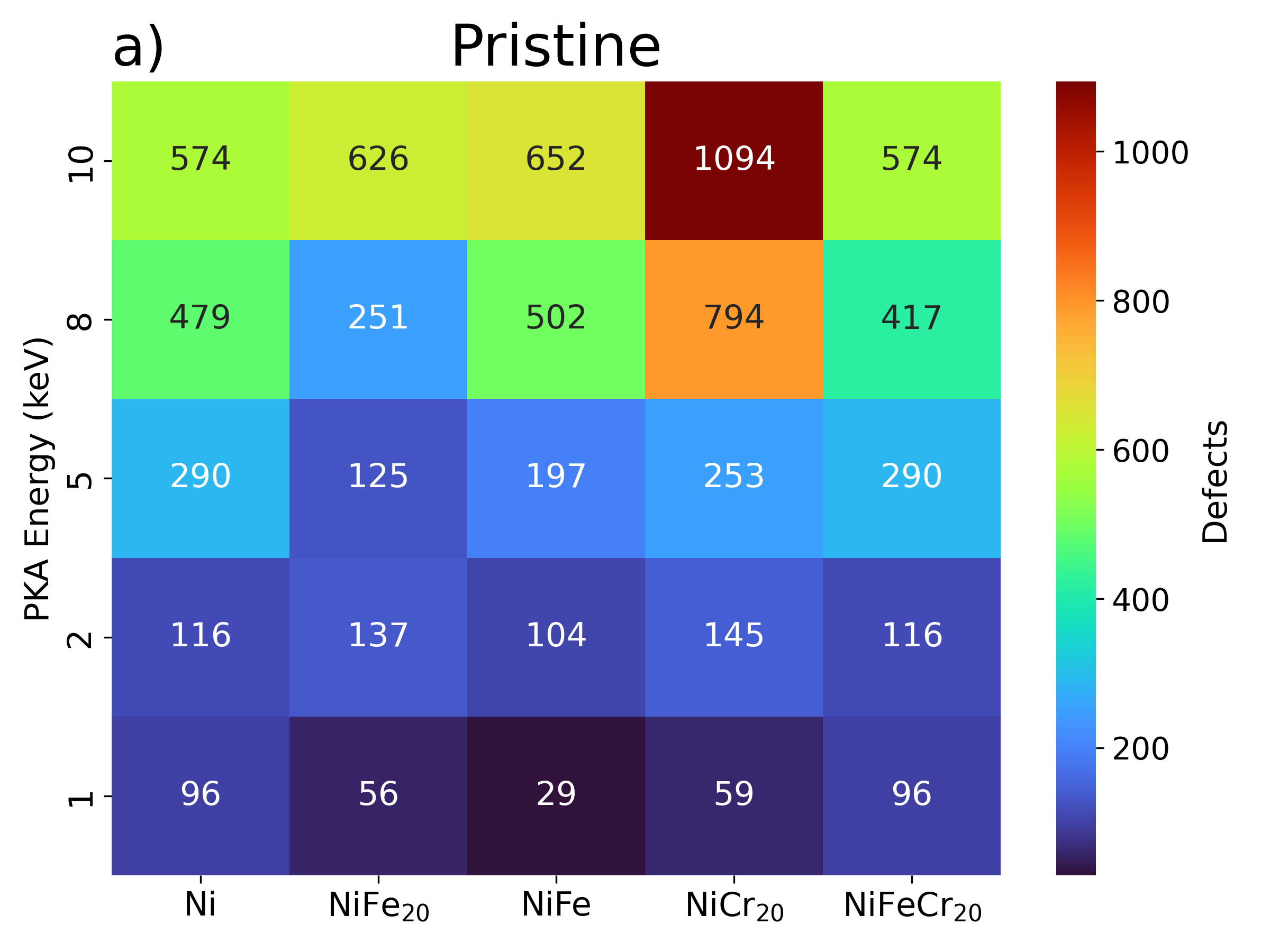}
    \includegraphics[width=0.40\textwidth]{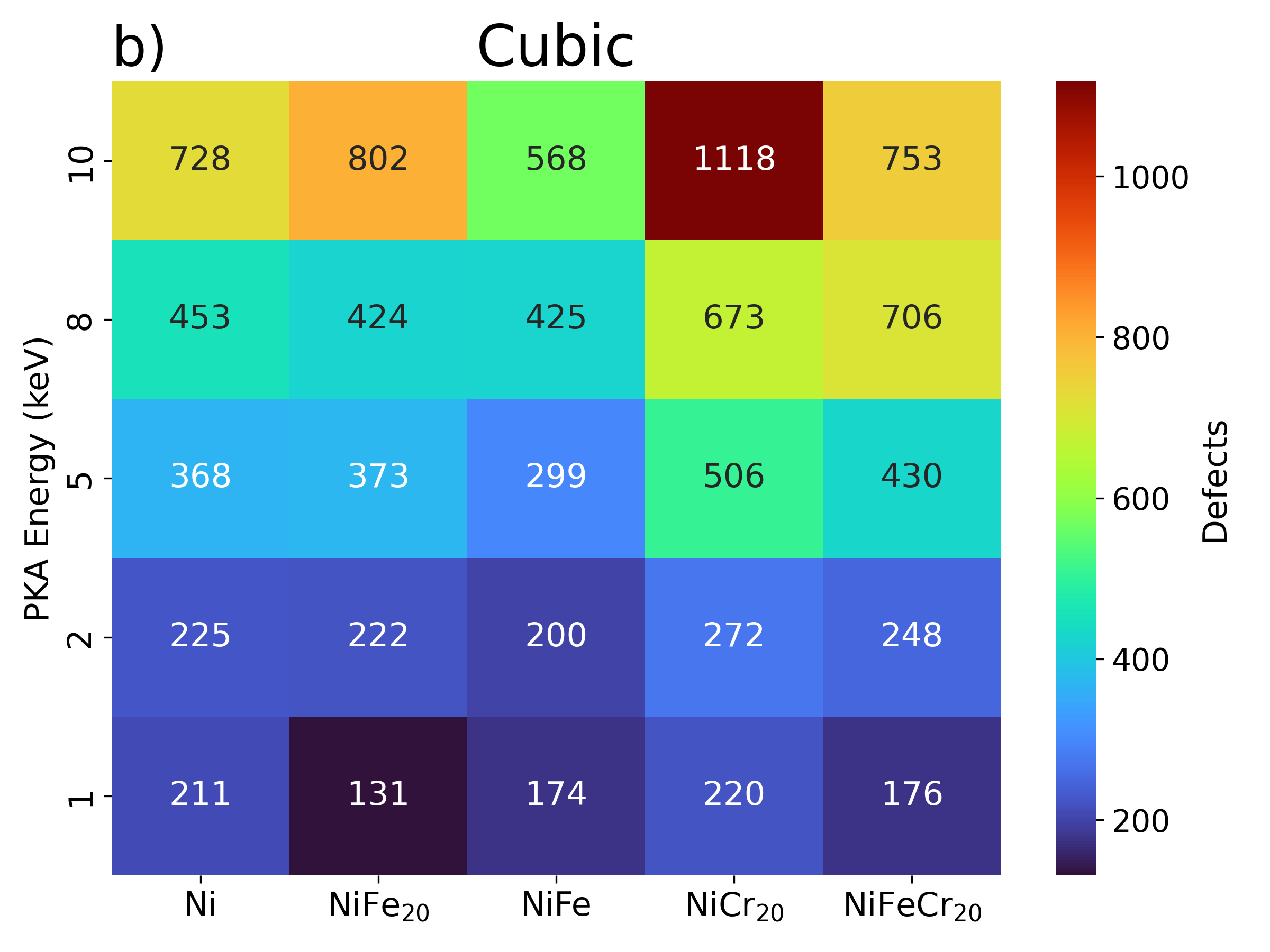}
    \includegraphics[width=0.40\textwidth]{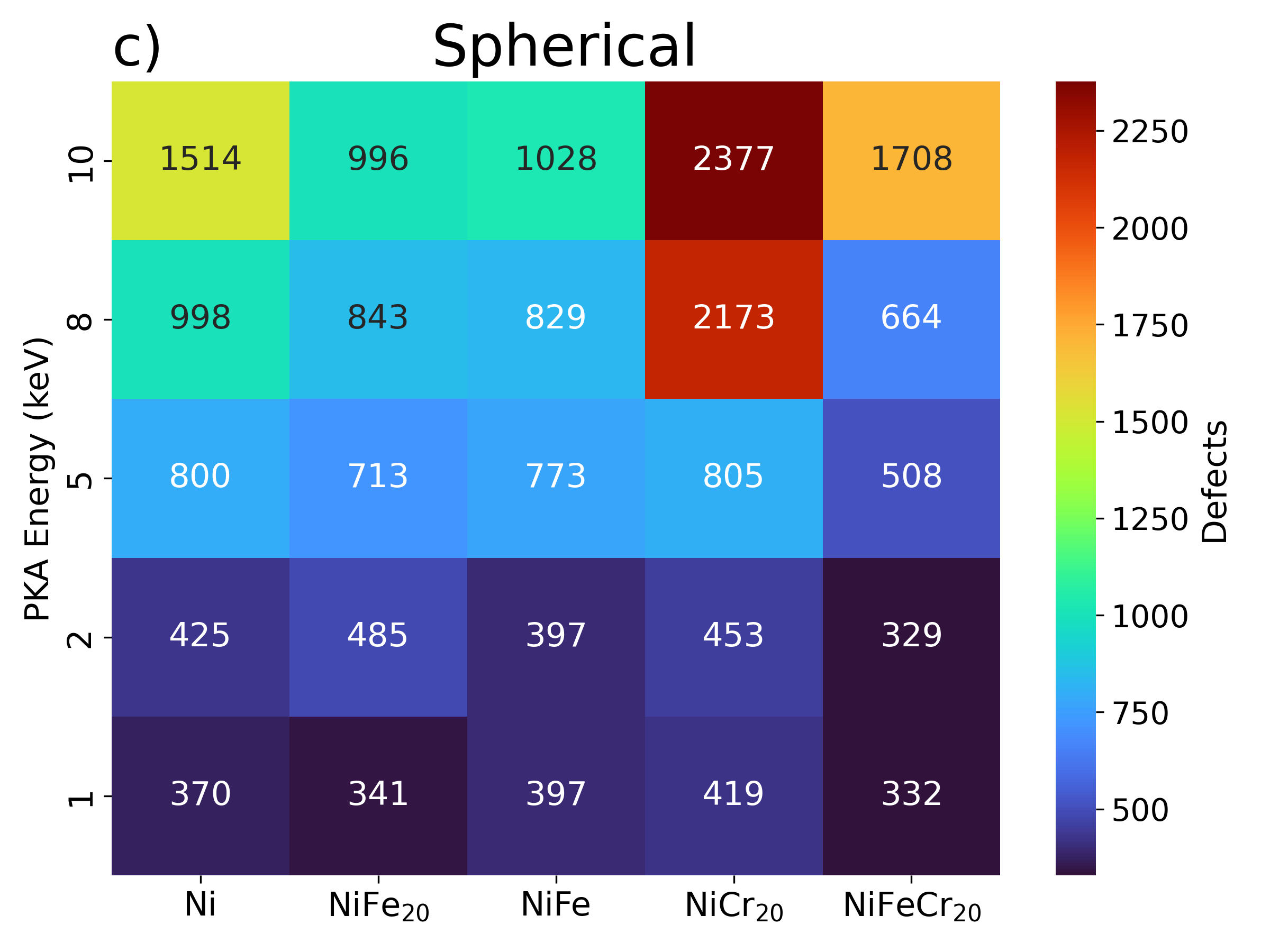}
    \caption{(Color online). Heat mapping displays the total number of
    defects after collision cascades for the pristine material in (b); 
    noticing the effect of Cr in the Ni matrix and NiFe alloys.}
    \label{fig:HeatMaps}
\end{figure}

In Fig. \ref{fig:HeatMaps}, the defects that are atoms with a 
different structure than FCC following collision cascades in all 
materials are illustrated in a heatmap graph within the PKA
energy range of 1 to 10 keV. Panel a) represents the pristine 
case, b) displays a cubic vacancy, and c) showcases a spherical 
void. It is evident that the presence of Fe atoms in the nickel 
sample reduces defect production compared to pure 
Ni, attributed to the soft bonding between Fe--Fe 
throughout the PKA range in the pristine and cubic vacancy 
scenarios. 
These results agree well with the results of the preliminary study, were we 
found that the formation energy of defects is increase with Fe 
concentration, Fig. \ref{fig:formation_energy}.
However, defect production increases for the spherical void due to the surface energy associated with this defect.
The reduction in defect production has been a subject of debate and may be
attributed to the potential \cite{beland2017accurate}. However, the studies
by Ullah et al. demonstrate that electron-phonon coupling induces defect
recovery and strain relaxation in NiFe alloys \cite{ullah2020electron}.
Conversely, the presence of Cr in the Ni sample increases
defect production due to Ni--Cr interactions, specially at 
high impact energies above 5 keV; regarless the shape of the preexisting 
defect or void. 
Of particular interest is the behavior of the NiFeCr$_{20}$ alloy
after collision cascades, where the total defect production is
comparable to pure Ni. This can be attributed to a compensation arising from 
the increase attributed to Cr solutes and the decrease due to Fe solutes.
%the interaction between Fe--Cr in the alloy, stabilizing the sample in 
%a manner similar to the mechanisms of material recovery observed in pure Ni.

\begin{figure}[b!]
    \centering
    \includegraphics[width=0.48\textwidth]{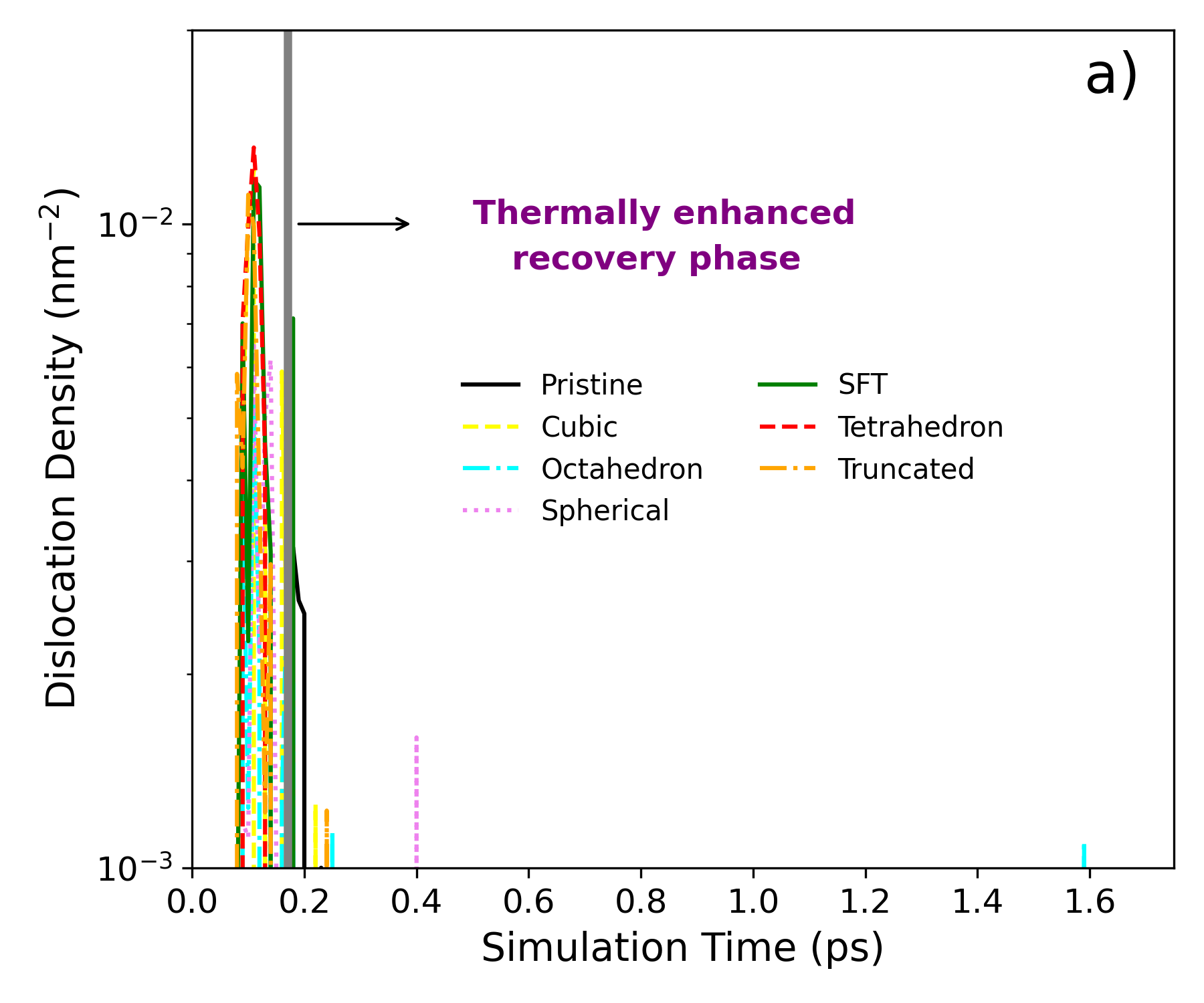}
    \includegraphics[width=0.48\textwidth]{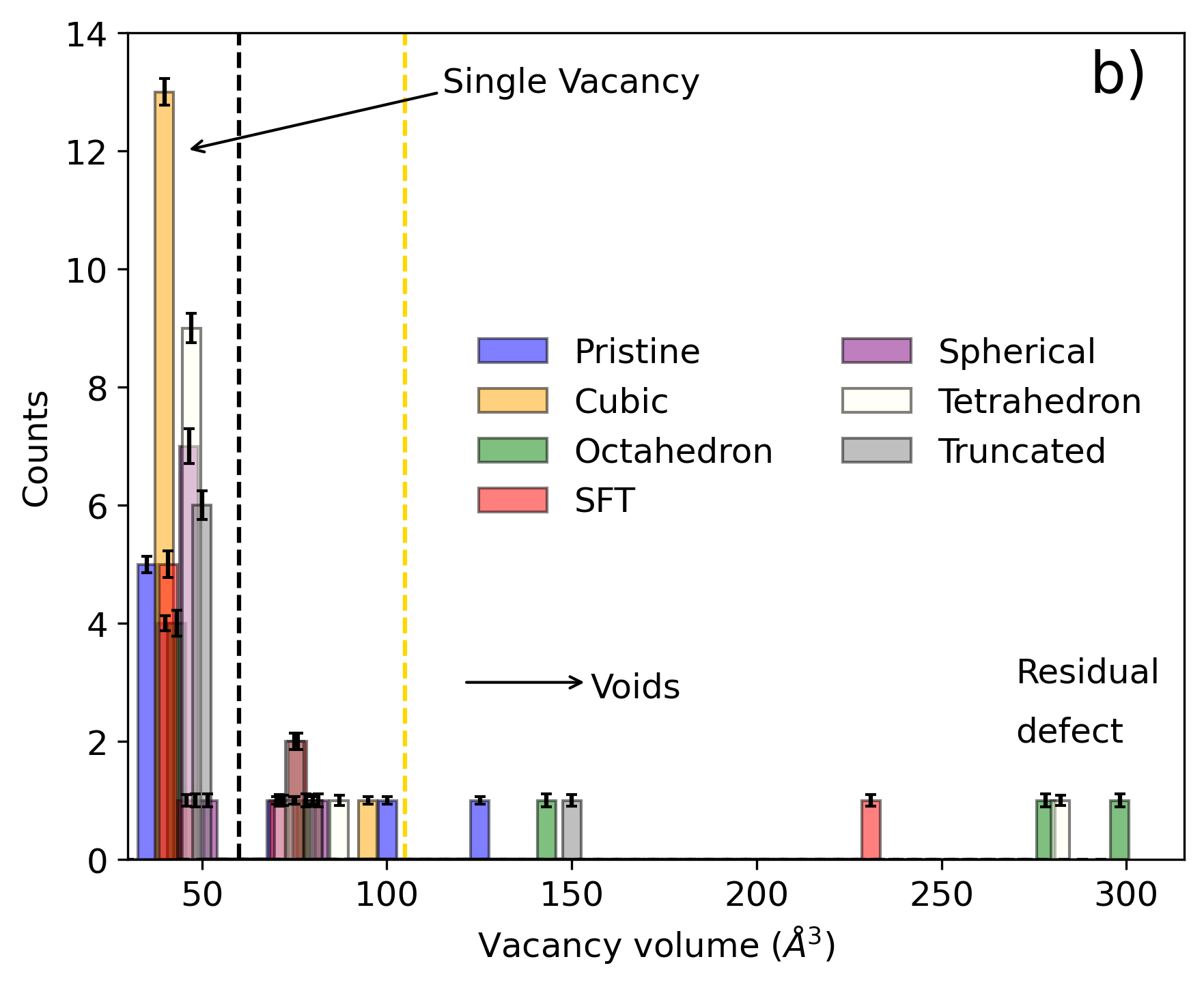}
    \caption{(Color online) Dislocation density evolution in NiCr$_{20}$ depicted over simulation time for total dislocations in (a), Shockley partials in (b), and stair-rod dislocations in (c) at a PKA energy of 8 keV. It is noteworthy that the nucleation of stair-rod junction dislocations and their stabilities are predominantly influenced by the presence of Shockley partials across various vacancy defects.}
    \label{fig:dislocNickelVoids}
\end{figure}

In Fig. \ref{fig:dislocNickelVoids}a), results depicting dislocation density as a function of simulation time for 
a pure nickel sample with 
various preexisting defects at 8 keV of PKA are presented. In Fig. 
\ref{fig:dislocNickelVoids}b), the quantification of single vacancies
and the clustering of vacancies to form voids is displayed for different
vacancy volumes. We observed that total dislocation nucleation occurs
around the heat spike and diminishes during the material's recovery,
regardless of the defects present in the samples before cascades.
However, the shape of the preexisting defect is crucial for the
formation of vacancies and voids. In a pristine case, only single 
with a volume of $\sim 35$ \AA{}$^{3}$ and
di/tri vacancies are created. An unstable cubic defect can produce
more single defects, and the most stable defect spherical void 
can only create single and di--vacancies. Finally, both a tetrahedron
and its truncated form demonstrate the ability to generate substantial 
voids following collision cascades. 
The former yields residual defects characterized by a significant
volume, akin to those produced by the octahedron defect.

%\begin{figure}[b!]
%    \centering
%    \includegraphics[width=0.48\textwidth]{Figures/FigNiFeCr20_defect_densities.png}
%    \caption{(Color online) Dislocation density evolution in NiFeCr$_{20}$ depicted over simulation time for total dislocations in (a), Shockley partials in (b), and stair-rod dislocations in (c) at a PKA energy of 10 keV. Legend is the same than in Fig. \ref{fig:dislocNiCr20time}.}
%    \label{fig:dislocationNiFeCr20time}
%\end{figure}

\subsection{The influence of Cr on dislocation nucleation.}

\begin{figure}[t!]
    \centering
    \includegraphics[width=0.48\textwidth]{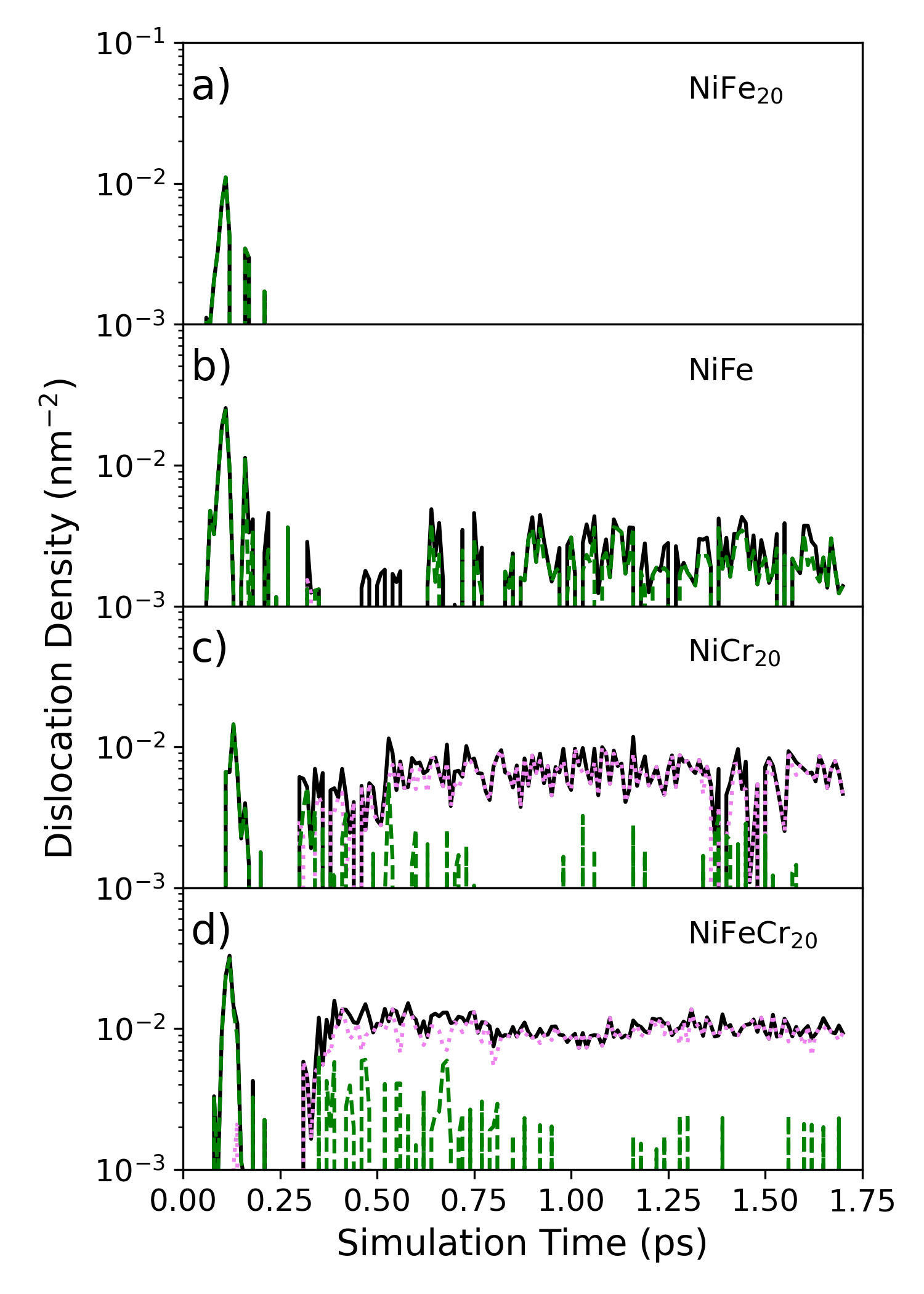}
    \caption{(Color online). Dislocation density as a function of simulation time for (a) pure NiFe$_{20}$, (b) NiFe, (c) NiCr$_{20}$, and (d) NiFeCr$_{20}$ alloys, considering a cubic defect with an 8 keV 
    Primary Knock-On Atom (PKA). The presence of Cr in either pure Ni and 
    NiFe alloys promotes the nucleation of Stair-Rod dislocation.
    \cite{fabris2018theoretical}.}
    \label{fig:dislocationCube}
\end{figure}

To analyze the effects of the presence of Fe and Cr in the Ni matrices,
Fig. \ref{fig:dislocationCube} illustrates the results for the
dislocation density over time in NiFe$_{20}$ in a), NiFe in b),
NiCr$_{20}$ in c), and NiFeCr$_{20}$ in d), considering a cubic defect
for total, Shockley partial, and Stair-rod dislocations.
When 20\% of Fe is in the Ni matrix, Shockley-type dislocations
are nucleated around the heat spike in a similar way to the observed
mechanisms for pure Ni. When the Fe concentration is increased to 
50\%, the Fe-Ni interaction and lattice mismatch make the Shockley
partial dislocations more stable during the material's recovery.
The presence of Cr in the Ni matrix is observed to impact the material's 
recovery, this is primarily due to lattice mismatch and interactions
between Ni-Cr, Cr-Fe, and Cr-Cr in the sample.

During the heat spike of the collision cascade, the nucleation of partial 
Shockley dislocations is observed for all samples.
Indeed, as the solutes induce a significant decrease in intrinsic stacking
fault energy, they encourage the nucleation of Shockley partials and a high density of stacking fault.
%{\color{blue}(Question: Twins in the samples ?)}
The interactions between different Shockley dislocations, as observed 
for the NiCr$_{20}$ and NiFeCr$_{20}$ alloys, can lead to the nucleation
of Stair--rod dislocations in some cases, as outlined below:
\begin{eqnarray}
    \frac{1}{6} [110] &=& \frac{1}{6} [\Bar{1} 2 1] + 
    \frac{1}{6} [2 \Bar{1} \Bar{1}] \quad \rm{Stair-rod},
\end{eqnarray}
and other symmetrical cases. This is illustrated in Fig. 
\ref{fig:dislocationCube}c-d), where the Stair-rod dislocation stabilizes,
and Shockley partials are annihilated after collision cascades 
a phenomenon further analyzed in subsequent sections of this
manuscript.

\begin{figure}[t!]
    \centering
    \includegraphics[width=0.48\textwidth]{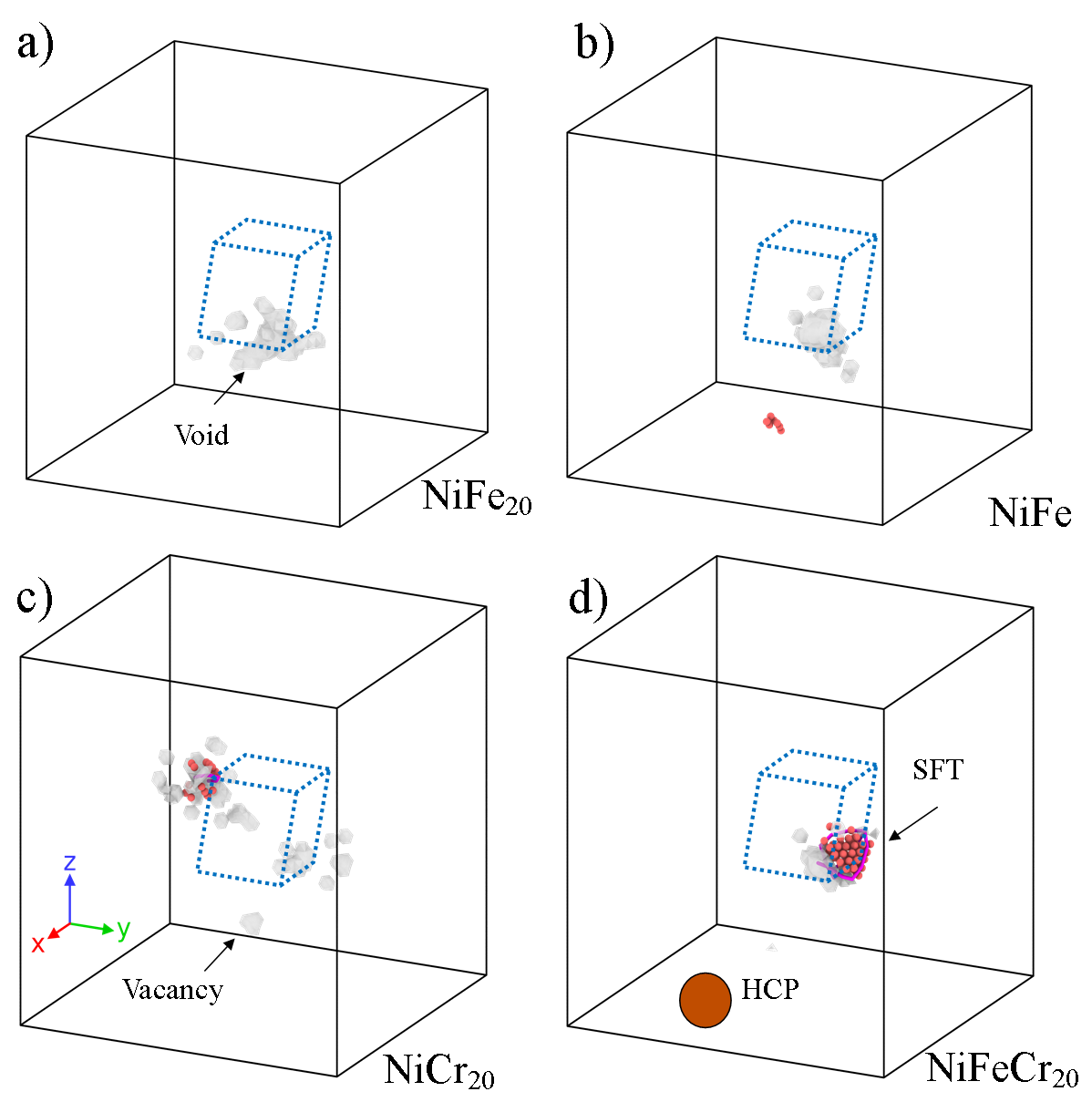}
    \caption{(Color online) Visualizations of vacancies, void formation, and 
    stacking fault tetrahedra (SFTs) identified by hexagonal close-packed (HCP) 
    atoms after collision cascades. In pure NiFe alloys in a) and b), 
    the distortion of the cubic defect is observed after collision cascades. 
    In NiCr$_{20}$ alloy in c), the influence of Cr atoms promotes stair-rod dislocation formation  and leads to SFT defects during material recovery. Similar effects are noted 
    in NiFeCr$_{20}$ in d), where the interaction between Fe and Cr stabilizes the 
    nucleation of SFT with stair-rod dislocation. }
    \label{fig:cubicVisualSFT}
\end{figure}

An interesting effect is observed when adding 20\% of Cr to
the pure Ni and NiFe alloy, where the interaction between Ni,Fe 
and Cr stabilizes the nucleation of SFT with a stair-rod dislocation during material
recovery, as depicted in Fig. \ref{fig:cubicVisualSFT}a--d).
The presence of Fe and Cr atoms in the Ni matrix stabilizes
the Shockley partials for certain defect vacancies,
and the mechanism of the interaction of Shockley partials
that nucleate a stair-rod junction persists.
It is noteworthy that the nucleation of Shockley partials
occurs during the material's recovery, where Ni and Fe
exhibit similar lattice constants, while Cr has a smaller
lattice constant. This lattice mismatch is further
accentuated for the NiFeCr$_{20}$ alloy, where the conversion
of one defect into another geometry is more prominently
observed. For instance, in the case of spherical vacancy
defects, they are transformed into three SFTs simultaneously.

\begin{figure}[t!]
    \centering
    \includegraphics[width=0.48\textwidth]{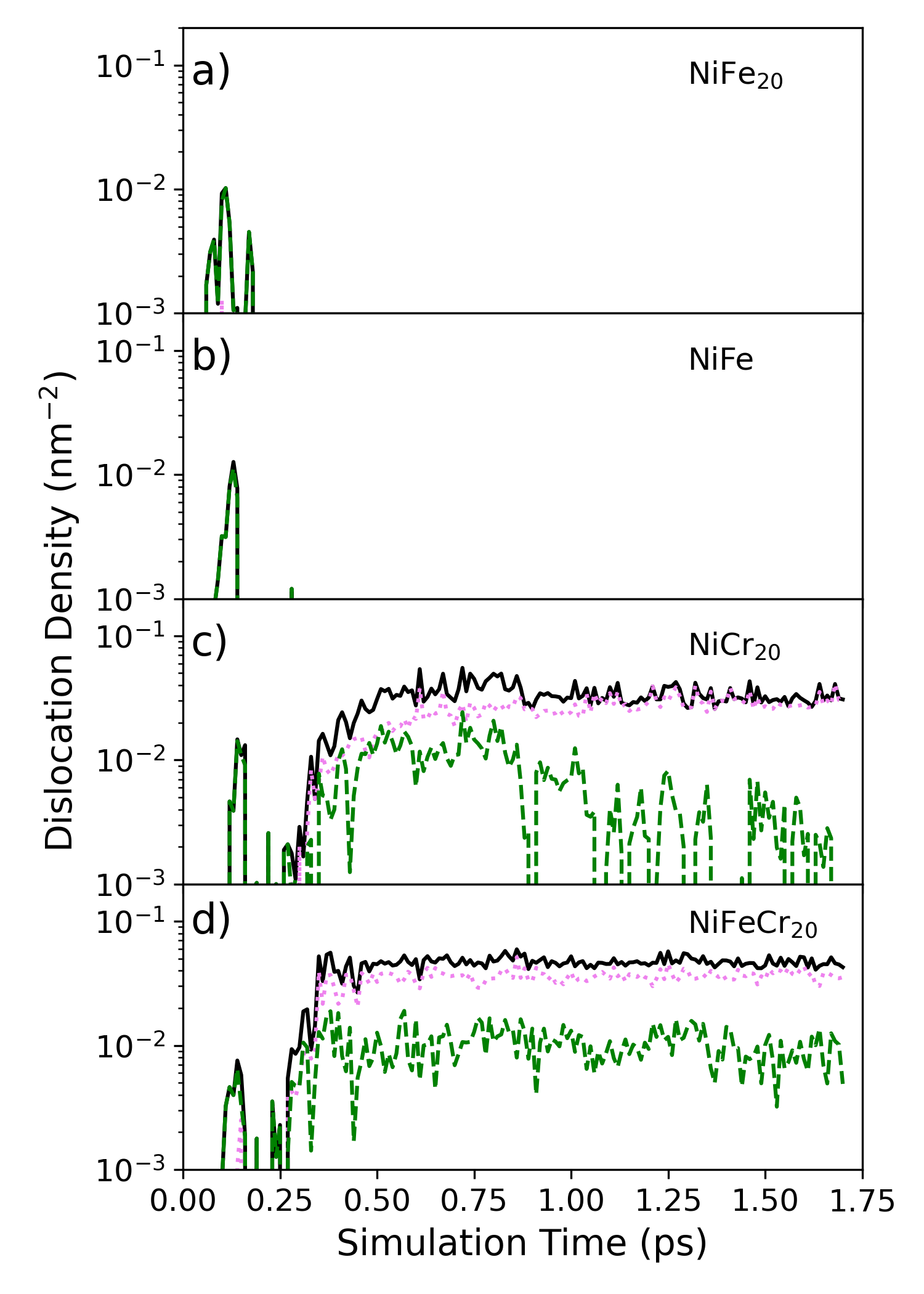}
    \caption{(Color online). Dislocation density profiles over time for 
    the spherical defect case at 8 keV PKA, illustrating the influence of 
    chemistry on pure Ni (a), NiCr$_{20}$ (b), and NiFeCr$_{20}$ (c). }
    \label{fig:structureSphere}
\end{figure}

In Fig. \ref{fig:structureSphere}, we present the dislocation density 
over simulation time for NiFe2$_{20}$ in (a), NiFe in (b), NiCr$_{20}$ in (c), 
and NiFeCr$_{20}$ in (d), considering a spherical void with 8 keV 
PKA.
Noticing that the formation of stacking fault tetrahedra (SFT) is more
pronounced in spherical defects within Cr-rich alloys. 
The chemical effects and the decrease of the intrisic stacking fault energy
on the NiFe CSAs are again observed through the
nucleation of stair--rod and Shockley dislocations during the material
recovery phase. 
Specifically, the NiCr$_{20}$ alloy nucleates a couple of SFTs
after collision cascades that is the alloy with the lowest stacking fault energy,
while the NiFeCr$_{20}$ alloy forms three SFTs where the spherical defect was initially located.
It is noteworthy that single collision cascades can transform unstable defects
into stable ones, influencing plastic deformation mechanisms during irradiation experiments. 
Importantly, the formation of SFT is exclusively observed for
these cubic and spherical defects among all considered
defect types. 
%{\color{blue} Question: In the NiCr20 and NiFeCr20 alloys without defects are all the shockley partial annihilate during the recovery ?}

\begin{figure}[t!]
    \centering
    \includegraphics[width=0.48\textwidth]{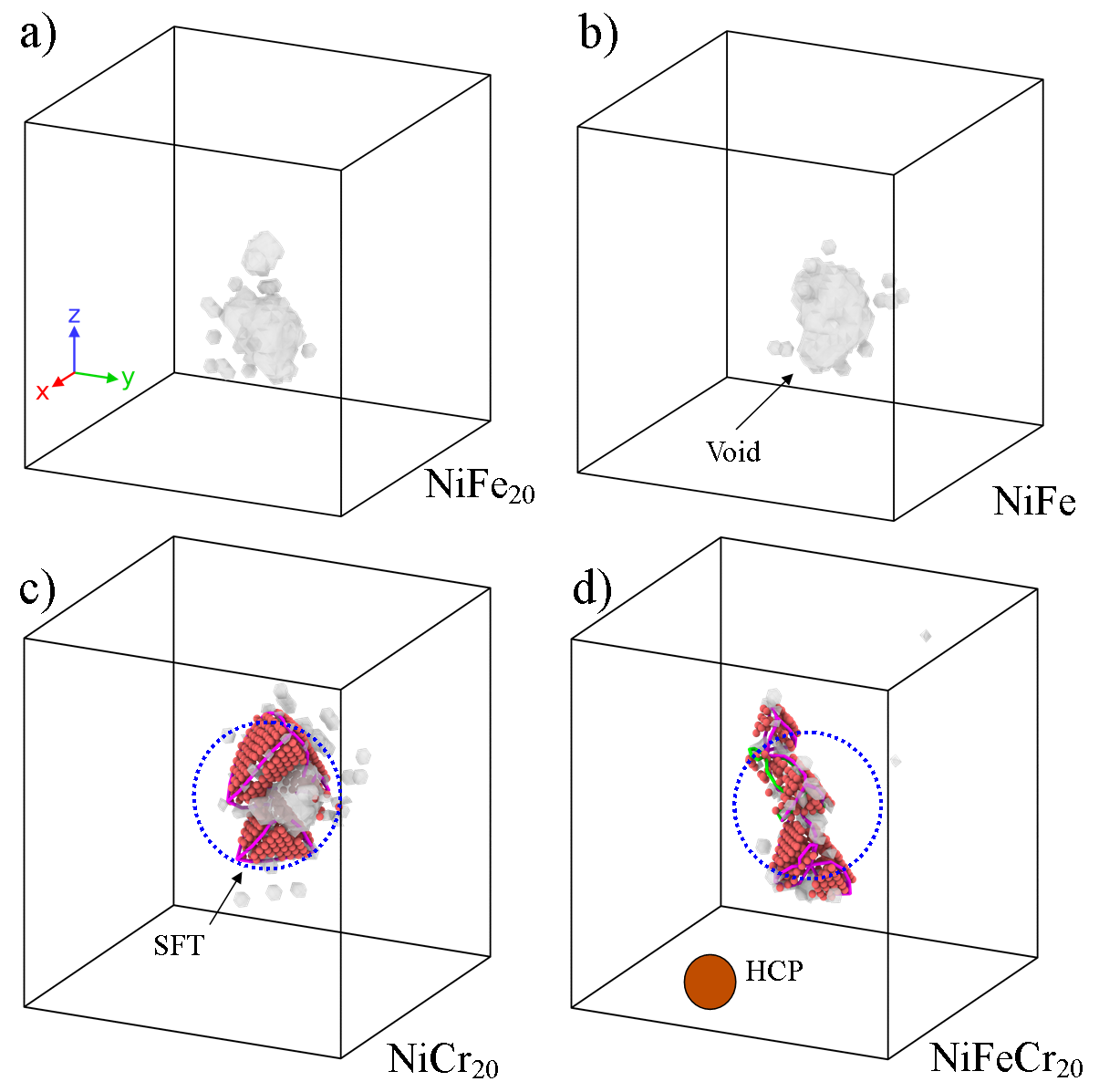}
    \caption{(Color online) The 
    recovery phase reveals nucleation of stair-rod and Shockley 
    dislocations, with NiCr$_{20}$ alloy forming a couple of stacking 
    fault tetrahedra (SFTs), and NiFeCr$_{20}$ alloy forming three SFTs 
    at the location of the initial spherical defect. Notably, SFT formation is exclusively observed for cubic 
    and spherical defects among all considered defect types.}
    \label{fig:SFT_void_sphere}
\end{figure}

In Fig. \ref{fig:SFT_void_sphere} shows the 
visualizations of spherical voids, vacancies, and stacking fault tetrahedra 
(SFTs) identified by hexagonal close-packed (HCP) atoms post-collision
cascades.
For NiFe CSAs, the formation of voids, vacancies, and some self-interstitial 
atoms (SIAs) is distributed throughout the sample deforming the defect without 
formation of SFTs.
In NiCr$_{20}$, the presence of Cr atoms has a significant impact on material 
recovery. Lattice mismatch and interactions between Cr-Cr and Cr-Ni promote
the generation of stair-rod dislocations, ultimately leading to the creation
of coupled SFTs defect post-collision cascades.

\begin{figure}[t!]
    \centering
    \includegraphics[width=0.48\textwidth]{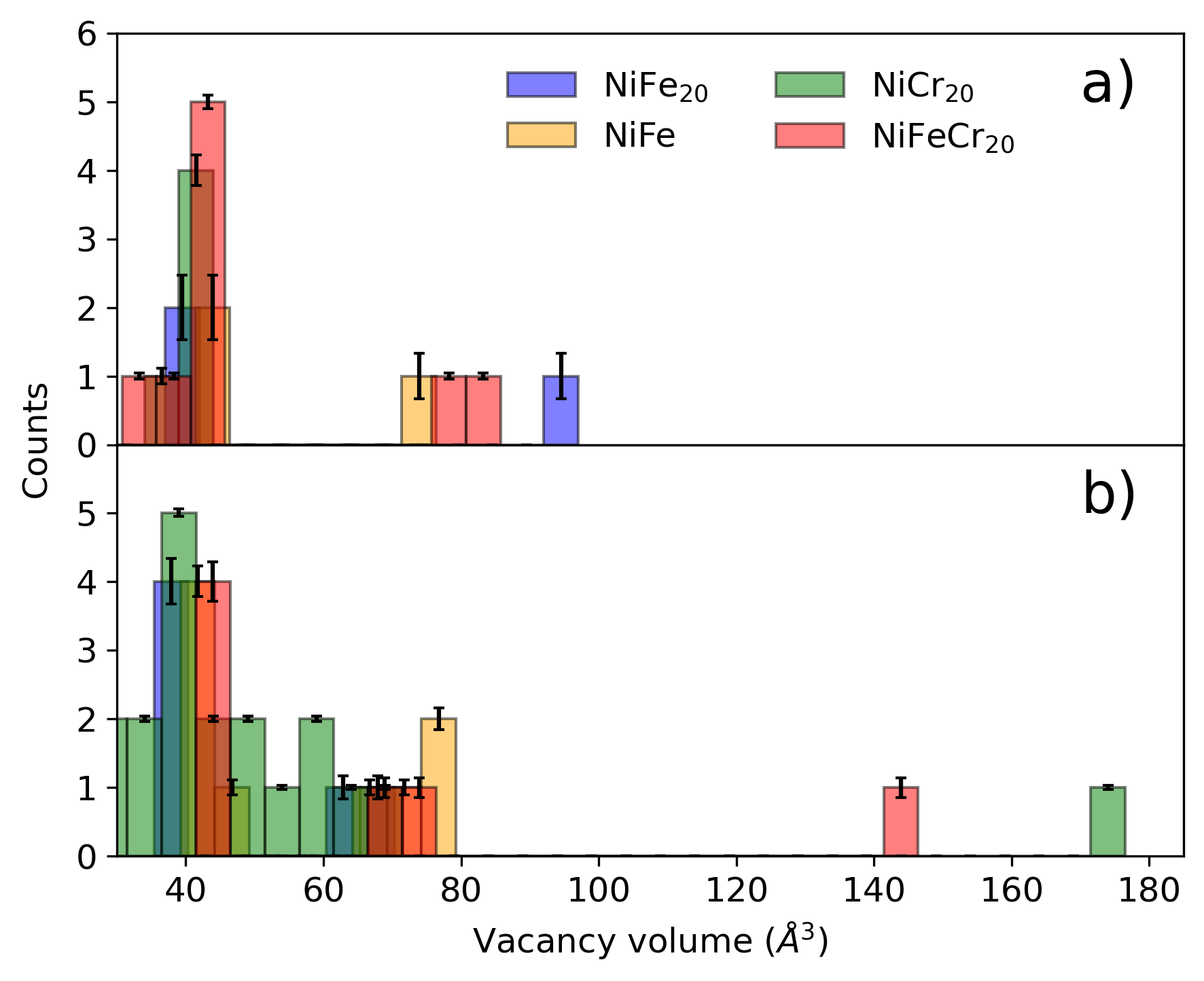}
    \caption{(Color online) Variation in vacancy counts as a function of their volume for different materials. Panel (a) illustrates the scenario with a cube defect, while panel (b) represents a spherical void, both simulated at 8 keV. It is evident that the presence of Cr in the Ni and NiFe CSAs significantly influences both defects, leading to the production of a greater number of single vacancies.}
    \label{fig:Void_vacancy_cube-sphere}
\end{figure}

In Fig. \ref{fig:Void_vacancy_cube-sphere} show the single vacancy and voids 
volume at the end of the collision cascade for cube defect in a) and a 
spherical void in b) at 8 keV. It is observed that NiFe CSAs for cubic are 
able to create sets of 3 vacancies, while Cr presence in Ni and NiFe CSAs 
tents to deform the cube defect into several single vacancies. 
For the spherical void, the NiFe CSAs deforms the void into sets of single 
and di vacancies, while Cr generates sets of single vacancies 
and a big void (roughly 6 vacancies) located at the center of the initial 
void. This efect is due to the GSF energy associated to the Cr-rich alloys 
as presented in Fig 1.

\section{Concluding remarks}
\label{sec:concl}

Summarizing, our computational exploration of single--phase
concentrated solid-solution alloys (SP-CSAs) has provided
valuable insights into the complex dynamics of
irradiation-induced defects. By conducting molecular dynamics
simulations across a spectrum of CSAs,
ranging from pure Ni, passing through binary alloys NiFe$_{20,50}$
and NiCr$_{20}$, 
to Ni$_{40}$Fe$_{40}$Cr$_{20}$, and varying PKA energies, we unraveled 
the interplay between alloying elements and defect evolution.
The modeling framework, considering a diverse array of vacancy
defects, illuminated a remarkable transition from spherical vacancy 
(voids) defects to stacking fault tetrahedra (SFTs). 
This transformation was intricately linked to the material's chemical 
composition, on the stacking fault energy of the alloys, 
by an analysis of the dislocation nucleation and evolution as 
a function of the dynamics time, with only Cr-rich alloys exhibiting
the nucleation of stair-rod dislocations, ultimately leading
to the formation of stable SFTs. Notably, NiCr$_{20}$ and NiFeCr$_{20}$ 
emerged as the exclusive materials capable of this intriguing mechanism. 
Indeed, both alloys exhibit very low stacking fault energies, 
facilitating the nucleation of SFTs validated by molecular static simulations 
to compute the surface, vacancy, and defect formation energies. 
%Highlighting that spherical voids are the most stable vacancy defects over all the considered cases after collision cascades.

These findings underscore the pivotal role of alloy chemistry in 
dictating irradiation-induced defect dynamics. The observed mechanisms, 
including the conversion to SFTs, hold implications for the irradiation 
resistance of SP-CSAs. This study contributes to a
understanding of the nanoscale phenomena governing vacancy 
defect evolution for informed material design 
strategies to enhance irradiation tolerance.

\section*{Acknowledgements}
%\ack
%We acknowledge 
Research was funded through the European Union Horizon 2020 research 
and innovation program under Grant Agreement No. 857470 and from the 
European Regional Development Fund under the program of the 
Foundation for Polish Science International Research Agenda PLUS, 
grant No. MAB PLUS/2018/8, and the initiative of the Ministry of 
Science and Higher Education 'Support for the activities of Centers 
of Excellence established in Poland under the Horizon 2020 program' 
under agreement No. MEiN/2023/DIR/3795. 
Computational resources were provided by the High Performance 
Cluster at the National Centre for Nuclear Research in Poland.
We would like to express our gratitude to NOETHER computing facilities 
at La Rochelle University and MCIA (Mésocentre de Calcul Intensif Atlantique).
Additionally, we extend our gratitude to GENCI - (CINES/CCRT), 
under Grant number A0110913037.
 %We acknowledge the computational resources 
 %provided by the Interdisciplinary Centre for Mathematical and
 %Computational Modelling (ICM) University of Warsaw under 
 %computational allocation no g91--1427.
 
\bibliographystyle{apsrev4-2}
\bibliography{biblio,references,bibliography}





% \end{thebibliography}
\end{document}